\documentstyle[10pt]{amsart}

\newcommand{\R}{{\mathbb R}}

\newcommand{\Z}{{\mathbb Z}}

\newtheorem{theo}{{\sc Theorem}}[section]
\newtheorem{defn}[theo]{{\sc Definition}}
\newtheorem{cor}[theo]{{\sc Corollary}}
\newtheorem{conj}[theo]{{\sc Conjecture}}
\newtheorem{lem}[theo]{{\sc Lemma}}
\newtheorem{prop}[theo]{{\sc Proposition}}
 
\newenvironment{proof}{\noindent{\em Proof:\/}}{\qed \medskip}

\title[uniformly bounded eigenfunctions]{Riemannian manifolds with uniformly bounded
eigenfunctions}

\author{John A. Toth and Steve Zelditch}
\address{Department of Mathematics and Statistics, McGill University,
Montreal, CANADA, H3A-2K6}
\address{Department of Mathematics, Johns Hopkins University, Baltimore, MD
21218, USA}

\thanks{\\ Research partially supported by NSERC grant \#OGP0170280\\ Research
partially supported by  NSF grant \#DMS-9703775}

\date{January 29, 2001}

\subjclass{58FO7, 58G25.}

\begin{document}

\maketitle

\addtolength{\baselineskip}{1pt}

\begin{abstract}
{\bf {\bf }}
The standard eigenfunctions $\phi_{\lambda} = e^{i \langle \lambda, x
\rangle}$
on flat tori $\R^n / L$ have $L^{\infty}$-norms bounded independently of the
eigenvalue.  In
the case of irrational flat tori, it follows that $L^2$-normalized
eigenfunctions have
uniformly bounded $L^{\infty}$-norms.  Similar bases exist on other flat
manifolds.
Does this property characterize flat manifolds? We give an affirmative
answer for
compact
Riemannian manifolds with completely integrable geodesic flows.
\end{abstract}

\setcounter{page}{1}
\setcounter{section}{-1}
\section{Introduction}

This paper is concerned with the  relation between the dynamics of the
geodesic flow
$G^t$ on the unit sphere bundle $S^*M$ of a compact Riemannian manifold
$(M,g)$ and
the growth rate of the $L^{\infty}$-norms of its $L^2$-normalized
$\Delta$-eigenfunctions (or `modes')
 $\{\phi_{\lambda}\}$.
   Let $V_{\lambda} := \{\phi: \Delta \phi_\lambda = \lambda \phi_\lambda\}$
denote the
$\lambda$-eigenspace for
$\lambda \in Sp (\Delta)$  and define
 \begin{equation} \label{L} L^{\infty}(\lambda, g) = \sup_{\phi\in V_{\lambda} \atop ||\phi||_{L^2}
= 1 }
 ||\phi||_{L^{\infty}},\;\;\;\;\;\;\;\ell^{\infty}(\lambda, g) = \inf_{ONB \{\phi_j\}
\in V_{\lambda}} (\sup_{j = 1, \dots, \dim V_{\lambda}} 
 ||\phi_j||_{L^{\infty}}). \end{equation}

The universal bound  $$L^{\infty}(\lambda, g) = 0(\lambda^{\frac{n-1}{4}})$$
holds for any $(M,g)$ in
 consequence of  the local Weyl law [Ho IV]
$$N(T, x): = \sum_{j: \lambda_j \leq T} |\phi_{\lambda_j}(x)|^2 =
\frac{1}{ (2\pi)^n} vol(M,g)
T^{\frac{n}{2}} + R(T,x),\;\;\;\;\;\;\;\;\;\;R(T,x) =
O(T^{\frac{n-1}{2}}).$$
It is attained in the case of the standard $(S^n, can)$ (by the  zonal
spherical harmonics) but is far off in the case of  irrational flat tori
$(T^n, ds^2)$
 where $L^{\infty}(\lambda, g) = O(1).$  These cases represent the extremes,
and the problem arises of characterizing the manifolds with extremal  growth
rates of $L^{\infty}$-norms of eigenfunctions.   In this article, we are  interested
in the case of  minimal growth:
\medskip

\begin{itemize}

\item {\bf Problem}: {\it Determine the $(M,g)$ for which
$\ell^{\infty}(\lambda,g) =
O(1)$ and those for which  
$L^{\infty}(\lambda,g) = O(1).$}\\

\end{itemize}

The same kind of problem may be posed in the more general setting of semi-classical Schroedinger operators $\hbar^2 \Delta + V$. 
The   eigenvalue problem $(\hbar^2 \Delta +V) \phi_{j} = E_j(\hbar) \phi_{j}$ now depends on $\hbar$, 
and we are interested in
the behaviour of eigenfunctions $\phi_{j}$ in the semiclassical limit  $\hbar \to 0$.  The spectrum becomes
dense around each regular value $E$ of the classical Hamiltonian $H(x, \xi)= |\xi|_g^2 + V(x)$ on $T^*M$, and for any $0 < \delta <1$,
the asymptotics of spectral data from an interval $[E - c \hbar^{1-\delta},  E + c \hbar^{1+\delta}]$ around $E$ will reflect the dynamics
of the classical Hamiltonian flow  $\Phi_t^{ E }$ on the  energy surface $X_{E }= \{H(x, \xi) = E \}.$   We fix $E$ and $0<\delta <1$, and consider the eigenvalues $ E_j(h) \in [E - c \hbar^{1-\delta} , E + c \hbar^{1-\delta} ].$
Denote by $V_{E_j(\hbar)}$ the eigenspace of eigenvalue $ E_j(h)$ and put: 
 \begin{equation}\label{SCL}  L^{\infty}(\hbar,E_j(\hbar) ; g, V ) = \sup_{\phi\in V_{E_j(\hbar)} \atop ||\phi||_{L^2}
= 1 }
 ||\phi||_{L^{\infty}},\;\;\;\;\;\;\; \ell^{\infty}(h, E_j(\hbar); g, V) =\inf_{ONB \{\phi_j\}
\in V_{E_j(\hbar)} } (\sup_{j = 1, \dots, \dim V_{E_j(\hbar)}} 
   ||\phi ||_{L^{\infty}}), \end{equation}  
and pose the analogous questions:
\begin{itemize}

\item {\bf Problem}: {\it Determine the $(M,g, V)$ for which there exists a regular  energy level $E$ such that \linebreak 
$\ell^{\infty}(\hbar, E_j(\hbar); g, V ) = O(1)$  and the $(M,g, V)$ for which $L^{\infty}(\hbar,E_j(\hbar); g, V) = O(1)$  as $\hbar \to 0$ with $E_j(\hbar) \in [E - c \hbar^{1 - \delta}, E+ c \hbar^{1 - \delta}]$ for some $c > 0$. } 

\end{itemize}

The problem on Laplace operators is the same as the problem on Schroedinger operators in the case   $V = 0$,
for any value of $E> 0.$

The problems about $\ell^{\infty}$   asks which Laplacians or Schroedinger operators possess an ONBE (orthonormal basis of
eigenfunctions)
of minimal growth.
The problems about $L^{\infty}$ ask which ones have the property that {\it every} ONBE has
minimal growth.  Obviously,
the distinction between $\ell^{\infty}$ and $L^{\infty}$   only arises when the spectrum of
$\Delta$ is
multiple.   At the opposite extreme, one may ask which $(M,g)$ possess
eigenfunctions which
achieve the maximal rate of growth, but we will not discuss that problem here.  One may also pose  quantitative problems
of  giving upper and lower bounds on 
$\ell^{\infty}(\lambda,g), L^{\infty}(\lambda,g)$, and their $L^p$-analgoues, 
under various dynamical hypotheses. Some results on such quantitative problems will be given in a subsequent article \cite{TZ}.

The known connections between  $\hbar^2 \Delta + V$- eigenfunctions and the dynamics of
$\Phi_t^E$
are not strong enough at present to answer these questions in the general
setting of compact
 Riemannian manifolds.   If, however,
the  systems are assumed to be  completely integrable geodesic flows then  much more can be
said.  

Let us assume in fact that $\Delta$ is {\it
quantum completely integrable}
in the (well-known) sense that there exist $P_1, \dots, P_n \in \Psi^1(M)$
($n$ = dim $M$)
 satisfying
\begin{itemize}

\item \;\; $[P_i, P_j]=0$;

\item \;\; $dp_1\wedge dp_2 \wedge \dots \wedge dp_n \not= 0$ on a dense
 open
 set  $\Omega \subset T^*M - 0$ of `finite complexity' (see below); 

\item \;\;
$\sqrt{\Delta} = \hat{K}(P_1, \dots, P_n)$ for some
polyhomogeneous function
$\hat{K}$ on $\R^n - 0.$
 \end{itemize}

Here, $\Psi^m(M)$ is the space of mth order pseudodifferential operators
over $M$, and
$p_k = \sigma_{P_k}$ is the principal symbol of $P_k$.  Since $\sigma_{[P_i,
P_j]} =
\{p_i, p_j\}$ (the Poisson bracket), it follows that the $p_j$'s generate a
homogeneous
 Hamiltonian action $\Phi_t$ of $t \in \R^n$ on $T^*M - 0$ with moment map
$${\cal P} : T^*M - 0 \rightarrow \R^n,\;\;\;\;\;\;{\cal P} = (p_1, \dots,
p_n).$$ We denote the image ${\cal P}(T^*M - 0)$ by $B$, by $B_{reg}$ (resp. $B_{sing}$)  the regular values
(resp. singular values)  of the moment map.

By `finite complexity' we mean the following: for each $b=(b^{(1)},...,b^{(n)})\in B$, let
$m_{cl}(b)$  denote the number of ${\Bbb R}^{n}$-orbits of the joint flow $\Phi_{t}$ on the level set ${\cal P}^{-1}(b)$.  Then
\begin{equation} \label{HYP} \mbox{\bf Finite complexity condition}:\;\;\;  \exists M: \;  m_{cl}(b) < M\;\;\;(\forall b \in B).  \;\;  \end{equation}
When $b \in B_{reg}$, then ${\cal P}^{-1}(b)$ is the union of $m_{cl}(b)$ isolated  Lagrangean tori. If $b \in B_{sing}$, then ${\cal P}^{-1}(b)$  consists of a finite number of connected
components, each of which is a finite union of orbits. These orbits may be Lagrangean tori, singular compact
tori (ie. compact tori of dimension $<n$) , or non-compact orbits consisting of cylinders or planes.

We will also  make the following assumption on the quantum level:
\begin{equation} \label{QHYP} \begin{array}{lllll} \mbox{\bf Bounded eigenvalue multiplicity }: & \exists M': \;m(\lambda) \leq M'\;\;  & &
 (\forall \lambda; \;m(\lambda) = dim V_{\lambda}). \end{array} \end{equation}
With this assumption,  $L^{\infty}$ is bounded by a constant times
 $\ell^{\infty}$, so all ONBE's are uniformly bounded if and only if one is. Without  assumption (\ref{QHYP}), it is simple to construct an
 ONBE which is not uniformly bounded.   We
will recall the construction in  \S 4, and discuss some open problems in which the
bounded eigenvalue multiplicity is dropped.

The Hamiltonian $|\xi|_g = \sqrt{\sum_{i,j = 1}^n g^{ij}(x) \xi_i \xi_j}$ is
then given
by $|\xi|_g  = K(p_1, \dots, p_n)$ where $K$ is the homogeneous term of
order 1 of
$\hat{K}.$    Hence the geodesic flow commutes with a Hamiltonian
$\R^n$-action, i.e. it is completely integrable.  We will assume throughout the following properness assumption:

Our main result is the following rigidity theorem: 
\medskip

\begin{theo} \label{RM}  Suppose that $\Delta$ is a quantum completely
integrable Laplacian on a compact Riemannian manifold $(M, g)$, and suppose that the  corresponding moment map satisfies (\ref{HYP}). Then:
\medskip

(a) If $L^{\infty}(\lambda, g) = O(1)$  then $(M,g)$ is flat.

(b) If $\ell^{\infty}(\lambda, g) = O(1)$ and if (\ref{QHYP}) holds, then $(M,g)$ is flat.
\medskip

More generally, suppose that $\hbar^2 \Delta + V$ is a quantum
completely
integrable Schroedinger operator,  and that the corresponding moment map ${\cal P}$  is proper and satisfies  (\ref{HYP}).   Assume there exists an energy level $E$ such that:
\medskip

(a)  $L^{\infty}(\hbar, E_j(\hbar));  g, V;) = O(1)$  as $\hbar \to 0$;

(b) $\ell^{\infty}(h, E_j(\hbar);  g, V) = O(1)$  as $\hbar \to 0$, and (\ref{QHYP})  holds.
\medskip

Then:  $E > \max V,$ and $(M,(E - V) g)$ is flat. If (a) (or (b)) holds for
all energy levels $E$ in an interval $E_1 < E < E_2$, then $(M, g)$ is flat and $V$
is constant.   

\end{theo}

As mentioned above, (a)-(b) are equivalent so we only consider (a) henceforth. 

We recall that flat manifolds are manifolds carrying a flat metric.  By
the Bieberbach theorems ([W], Theorems 3.3.1 - 3.3.2),
 a flat manifold $(M, g)$ may be expressed as the quotient
$M= \R^n/\Gamma$ of $\R^n$ by a discrete (crystallographic) subgroup of
Euclidean motions $\Gamma \subset E(n).$ The subgroup $\Gamma^* :=\Gamma
\cap \R^n$ is normal and of finite index
in $\Gamma$ so there exists a flat torus $T^n = \R^n /\Gamma^*$ and a finite
normal Riemannian
cover $\pi: T^n \to M$ with deck transformation group $G = \Gamma/\Gamma^*.$
For each $n > 0$, there are only
finitely many affine equivalence classes of flat compact connected $(M,g)$
of dimension $n$ (affinely equivalent=
same fundamental group), and in  low dimensions they have been classified
(cf. [W]).  The eigenfunctions $\phi_{\lambda}$ of
$\Delta_g$ on
$(M, g)$ may be lifted to $G$-invariant eigenfunctions $\pi^*
\phi_{\lambda}$ on $T^n$
and hence the eigenspace $E_{\lambda}(M,g)$ may be identified with the
$G$-invariant
eigenspace $E_{\lambda}(T^n, g_T)^G$.  The latter eigenfunctions may be
written as
sums of exponential functions.

Let us  outline the proof of the Theorem  (\ref{RM}) in the simplest case of
 {\it toric } integrable systems (see \S 1 for background), and then explain what more is involved in the case of general integrable systems.
 By definition, the geodesic flow $G^t_g: T^*M \rightarrow T^*M$ of
a compact Riemannian manifold $(M,g)$ is  toric integrable if it
 commutes with a Hamiltonian action of the n-torus $\R^n/\Z^n$.
  Equivalently, if  there exist global action variables $\{(I_j, \theta_j): j=1, \dots, n\}$
 for the geodesic flow, i.e. functions of $(p_1, \dots, p_n)$ whose Hamilton
flows are
$2\pi$-periodic.  The level sets
$$T_I :=  {\cal I}^{-1} (I)$$
 of the moment map
$${\cal I}= (I_1, \dots, I_n): T^*M-0 \rightarrow \R^n$$
are then orbits $\R^n/\Z^n \cdot (x_o, \xi_o)$ of the torus action and hence
are tori. The image $B$ of $T^*M - 0$ under  ${\cal I}$ is a convex polyhedral cone
and ${\cal I}$ is a Lagrangean torus bundle over its interior.
Such moment maps ${\mathcal I}$ are the cotangent bundle analogues of toric varieties in algebraic
geometry.

In the toric case, it is always possible to quantize the action variables as first
 order pseudodifferential {\it  action operators} $\hat{I}_j$
which commute with $\Delta$.  The actions define a (projective) action of
$\R^n/\Z^n$
by Fourier integral operators, or equivalently, the joint spectrum
$Sp(\hat{I}_1, \dots,
\hat{I}_1)$ is contained in an (off-centered) lattice $\Z^n + \mu$.
The joint
eigenfunctions
$$(\hat{I}_1,\dots,\hat{I}_n) \phi_{\lambda} = \lambda \phi_{\lambda} \;\;\;
\lambda \in \R^n$$
are therefore  quantizations of  the  invariant Lagrangean torii ${\cal
T}_{\lambda}$
with integral actions $\lambda \in \Z^n + \mu$.  In particular,
eigenfunctions
$\{\phi_{\lambda}\}$ {\it localize on the invariant tori} in the semiclassical limit
in the sense that for
any zeroth order pseudodifferential operator $A$ (with symbol $\sigma_A)$,
\begin{equation}\label{LOC}  (A \phi_{k \lambda}, \phi_{k \lambda})    =
\int_{T_{\lambda}} \sigma_A d\mu_{\lambda} +
O(k^{-1}),\end{equation}
where $d\mu_{\lambda}$ is the normalized Lebesgue (probability) measure on
$T_{\lambda}.$
Hence,
 $| \phi_{\lambda}(x)|^2$ measures the density of
the natural  projection  $\pi_{\lambda}: {\cal T}_{\lambda} \rightarrow M$
at $x$. 

The proof of Theorem (\ref{RM}) in the toric case is based on the following simple Lemmas.  First
we have:
\medskip

{\it  Suppose that $G^t$ is toric
integrable and  that $L^{\infty} (M,g) = 0(1)$. Then  every
invariant torus $T_{\lambda}$ has a non-singular projection to $M$.}
\medskip

The proof uses the fact that for any invariant torus $T_I$, there exists a
sequence of  joint eigenfunctions $\{\phi_{\lambda}\}$ of the quantum torus
action which localizes on $T_I$.
Uniform boundedness of the eigenfunctions then implies regular projection of
the tori. 

The second ingredient in the proof of the main theorem  in the case of toric
integrable systems
 is the following
 purely geometric statement which follows from the recently proved Hopf
conjecture
(cf. \cite{BI} \cite{CK}).
\medskip

{\it Suppose that $(M,g)$ is a compact Riemannian
manifold with
toric integrable geodesic flow, and suppose that all the invariant torii
project regularly to $M$.  Then $(M,g)$ is a flat manifold.}
\medskip

By `projecting regularly' we mean that the projection has no singular values,
hence (in view of the
dimensions) is a  covering map.

The proof of Theorem (\ref{RM}) in the case of
general  Hamiltonian $\R^n$ actions is basically similar, but there are some
new complications to handle.
Geometrically, the new features are that the fibers ${\mathcal P}^{-1}(b)$
may have several components (`geometric multiplicity'),  that there may exist
non-compact orbits (e.g. embedded
cylinders), and that there may exist singular orbits lying over the interior of
the image of $T^*M-0$ under ${\mathcal P}.$
Analytically, the main new feature is  that modes need not localize on
individual components of  ${\mathcal P}^{-1}(b)$.
What does localize on individual tori are {\it quasimodes}, i.e. semiclassical
Lagrangean distributions which  approximately solve the eigenvalue problem.  In
the toric case, modes and quasimodes are the same but this is not the case in general.
As originally stressed by Arnold \cite{A},
and as is evident from simple examples such as the symmetric double well
potential, eigenfunctions may be linear
combinations of quasi-modes with very close quasi-eigenvalues and in the
classical limit their mass concentrates
in some way on the union of the components.  How the mass is distributed involves
the question whether the tori are {\it resonant} or not, and whether or not there
is tunnelling between tori. We will discuss such  relations between modes and quasimodes
 in detail in \cite{TZ}, where we prove (among other things) that quasimodes have uniformly
bounded sup norms when modes do and where we determine precisely how modes blow up around singular orbits.
In this paper, we take a softer approach via quantum limits of eigenfunctions and semiclassical
trace formulae.

We close with some acknowledgements. We thank Bruce Kleiner for pointing out the paper \cite{M}, Leonid Polterovich for helpful comments on  \cite{BP}, 
and Francois Lalonde for helpful comments on an earlier version of the
paper.
 We would especially like to thank the referee of this paper for pointing
out that one of our original (non-degeneracy) hypotheses could be removed from the proof of Theorem (\ref{RM}),
and for several other  corrections and improvements. To clarify the ingredients in the proof,  we cut the original manuscript (which appeared on the lanl archive
as  math-ph/0002038) into two parts, the present qualitative one and the subsequent quantitative one (\cite{TZ}).

\section{Background}

\subsection{Completely integrable systems}

By a completely integrable system on $T^{*}M$ we mean  a set of
$n$ independent,   $C^{\infty}$ functions  $p_1, \dots, p_n$, 
 on $T^{*}M$ satisfying:

\medskip

\begin{tabular}{l} $\bullet \,\, \{ p_{i}, p_{j} \}=0$ \,\, for all $1 \leq i,j \leq n$;\\
$\bullet \,\, dp_{1} \wedge dp_{2} \wedge \cdot \cdot \cdot \wedge dp_{n} \neq 0$ \,\, on an open dense subset of $T^{*}M.$
\end{tabular}
\medskip

The associated moment map is defined by
\begin{equation} \label{MM} {\cal P} = (p_1, \dots, p_n): T^*M \rightarrow
B \subset\R^n. \end{equation}  We  refer to to the set $B$ as the `image of the moment map.' 
The Hamiltonians generate an action of $\R^n$ defined by 
$$ \Phi_t = \exp t_1 \Xi_{p_1} \circ \exp t_2 \Xi_{p_2} \dots \circ \exp t_n
\Xi_{p_n}.$$
We often denote  $\Phi_t$-orbits by   $\R^n \cdot (x, \xi) $. The isotropy group of $(x, \xi)$ will be denoted by 
 ${\mathcal I}_{(x, \xi)}.$ When  $\R^n \cdot (x, \xi) $ is a compact Lagrangean orbit, then ${\mathcal I}_{(x, \xi)}$
is a lattice of full rank in $\R^n$, and is  known
as the `period lattice', since it consists of the `times' $T \in \R^n$ such that $\Phi_T |_{\Lambda^{(j)}(b)} = Id.$

We will need the following:
\begin{defn}\label{SING}
We say that:

\begin{itemize}

\item  $b \in B_{sing}$ if ${\cal P}^{-1}(b)$ is a singular level of the moment map, i.e. if
there exists a point $(x, \xi) \in {\cal P}^{-1}(b)$  with  $dp_{1} \wedge \cdot \cdot \cdot \wedge dp_{n}(x,\xi) = 0$.
Such a point $(x, \xi)$ is called a singular point of ${\cal P}$.

\item a connected component of ${\cal P}^{-1}(b)$ ($b \in B_{sing}$)is a singular component if it  contains a singular point ;

\item an orbit $\R^n \cdot (x, \xi) $ of $\Phi_{t}$ is  singular if it is non-Lagrangean, i.e.   has dimension $<n$;

\item $b \in B_{reg}$ and that ${\cal P}^{-1}(b)$ is a regular level if all points $(x, \xi) \in {\cal P}^{-1}(b)$ are regular, i.e. if $dp_{1} \wedge \cdot \cdot \cdot \wedge dp_{n}(x, \xi) \not= 0$.

\item   a component of  ${\cal P}^{-1}(b)$ ( $b \in B_{sing} \cup B_{reg}$) is regular if it contains no singular points.

\end{itemize}

\end{defn}

By the Liouville-Arnold theorem [AM], the orbits of the joint flow $\Phi_{t}$ are
diffeomorphic to $\R^k \times T^m$ for some $(k,m), k + m \leq n.$
By the properness assumption on  ${\cal P}$, a regular level  has the form
\begin{equation} \label{CI1}
{\cal P}^{-1}(b) = \Lambda^{(1)}(b) \cup \cdot \cdot \cdot \cup
 \Lambda^{(m_{cl})}(b) , \;\;\;(b \in B_{reg})
\end{equation}
\noindent where each $\Lambda^{(l)}(b) \simeq T^n$ is an $n$-dimensional Lagrangian
torus.  The  classical (or geometric) multiplicity function   $m_{cl} (b) = \# {\cal P}^{-1} (b)$,
i.e. the number
of  orbits on the level set ${\cal P}^{-1}(b)$, is constant on    connected components of $B_{reg}$ and the moment map
(\ref{MM}) is a fibration over each component with fiber (\ref{CI1}).
 In  sufficiently small neighbourhoods $\Omega^{(l)}(b)$ of each component torus,
$\Lambda^{(l)}(b)$, the Liouville-Arnold theorem also gives the existence of local action-angle variables 
$(I^{(l)}_{1},...,I^{(l)}_{n}, \theta^{(l)}_{1},...,\theta^{(l)}_{n})$ in terms
of which the joint flow of $\Xi_{p_{1}},...,\Xi_{p_{n}}$ is linearized [AM]. For convenience, we henceforth normalize the action variables $I^{(l)}_{1},...,I^{(l)}_{n}$ so that $I^{(l)}_{j} = 0; \, j=1,...,n$ on the torus $\Lambda^{(l)}(b)$.

When $b \in B_{reg}$, the Lagrangean tori $\Lambda^{(j)}(b)$ of ${\cal P}^{-1}(b)$ carry two natural measures,
which we take some care to distinguish.

\begin{defn} We define:
\begin{itemize}

\item Lebesgue measure $d\mu_b^{(j)} = (2\pi)^{-n} d \theta_1 \wedge \cdots \wedge d\theta_n $  on $\Lambda^{(j)}(b)$, as the  normalized (mass one) $\Phi_t$-invariant measure
on this orbit;

\item  The Liouville measure $d\omega^{(j)}_b$ on  $\Lambda^{(j)}(b)$, as the surface measure induced by the
moment map ${\mathcal P}$, i.e. 
$$d\omega^{(j)}_b = \frac{dV}{dp_1 \wedge \cdots \wedge dp_n}$$
where $dV$ is the symplectic volume measure on $T^*M.$  
By the Liouville mass of $\Lambda^{(j)}(b)$  we mean
the integral 
$$ \omega^{(j)}(b):= \int_{\Lambda^{(j)}(b)}  d\omega^{(j)}_b . $$
 \end{itemize}

\end{defn}

 The Liouville mass of a compact Lagrangean orbit $\Lambda^{(j)}(b)$  has a simple dynamical interpretation: 
it is the Euclidean volume of the fundamental domain of the common period lattice ${\mathcal I}^{(j)}_b = 
{\mathcal I}^{(j)}_{(x, \xi)}$
of points $(x, \xi) \in \Lambda^{(j)}(b)$  , i.e.
\begin{equation} \label{LM}\omega^{(j)}(b) = Vol(\R^n/ {\mathcal I}^{(j)}_b). \end{equation}
Indeed, by writing Liouville measure in local action-angle variables, we see that
\begin{equation} d\omega^{(j)}_b = \det (T^k_{\ell}(b)) d\mu_b^{(j)}, \;\;\;\mbox{where}\;\; T^{k}_{\ell} = \frac{\partial I_k}
{\partial p_{\ell}}. \end{equation}
It is clear from the definition of the action-angle variables that $ {\mathcal I}^{(j)}_b$ is generated by the
rows $(T^k_1, \dots, T^k_n)$, hence the determinant is the co-volume of the period lattice.

We now turn to singular levels.
When  $b \in B_{sing}$ we first decompose 
\begin{equation} \label{SL} {\cal P}^{-1}(b) = \cup_{j = 1}^{r} \Gamma_{sing}^{(j)}(b) \end{equation}
the singular level into connected components $\Gamma_{sing}^{(j)}(b)$ and then decompose
\begin{equation} \label{ORB} \Gamma_{sing}^{(j)}(b) = \cup_{k = 1}^{p} \R^n \cdot (x_k, \xi_k) \end{equation}
each component  into orbits. Both decompositions can take a variety of forms.  The regular components $\Gamma_{sing}^{(j)}(b)$ must be Lagrangean tori by the properness assumption. A singular components consists of finitely many orbits by
the finite complexity assumption. The orbit $\R^n \cdot (x, \xi)$ of a singular point is necessarily singular,
hence has the form
$\R^k \times T^m$ for some $(k,m)$ with  $k + m < n.$ Regular points may also occur on a singular component, whose
orbits are Lagrangean and can take any one of the forms $\R^k \times T^m$ for some $(k,m)$ with  $k + m = n.$

We will need the following result in the proof of Theorem (\ref{RM}):

\begin{prop}\label{EXISTS} A singular component $\Gamma_{sing}^{(j)}(b) \subset {\cal P}^{-1}(b)$ (with $b \in B_{sing}$)
must contain a compact singular orbit $\R^n \cdot (x, \xi) \simeq T^k, k < n.$ \end{prop}

\begin{proof}
It follows by a standard averaging argument \cite{M2} that
the set ${\mathcal M}_{\Gamma_{sing}^{(j)}}^I$ of invariant probability measures  supported on  $\Gamma_{sing}^{(j)}$
is non-empty: for any probability measure $\mu_0$ supported on $\Gamma_{sing}^{(j)}$, the set of weak* limit
points of the set of  finite  time
averages $\mu_T = \frac{1}{vol \{|t| \leq T\}} \int_{|t| \leq T} (\Phi_t)_* \mu_0 \, dt$ gives at least one non-trivial
element of ${\mathcal M}_{\Gamma_{sing}^{(j)}}^I$.  Since $\Gamma_{sing}^{(j)}$ consists of only finitely many
orbits, any invariant measure in  ${\mathcal M}_{\Gamma_{sing}^{(j)}}^I$ is a finite sum of (ergodic) measures,
each supported on just one orbit. The non-compact  orbits
$\R^k \times T^m$  obviously cannot carry
 invariant probability measures; hence, at least one orbit must be compact.
\end{proof}

We will need a further result on Hamiltonian $\R^n$-actions $\Phi_t$. We define a non-zero period of $\Phi_t$ to
be a time $T \in {\mathcal I}^{(j)}_b - \{0\}$ for some $(b, j),$ and denote the set of periods by ${\mathcal T}$.

\begin{prop} \label{YORKE} There exists  a constant $C > 0,$ which depends on the Riemannian manifold $(M,g)$, such that $\inf_{\{ T \in {\mathcal T} \}} |T| \geq C.$ \end{prop}

\begin{proof} In the case of a Hamiltonian flow  with Hamilton vector field $\Xi$, this is a case of  Yorke's theorem \cite{Y}.  In fact,
$C = \frac{2\pi}{L}$ where $L = || d \Xi||_{\infty}$.  In the case of $\R^n$ actions, we can apply Yorke's theorem
to any one parameter subgroup. \end{proof}

\subsection{   Hamiltonian torus actions}

In special cases  (see  \cite{D} for the geometric conditions), the Hamiltonian
$\R^n$ action
descends to the  Hamiltonian  action of the torus $\R^n/\Z^n$ on $T^*M$.  Such  Hamiltonian torus actions are the cotangent space analogues of toric varieties in algebraic geometry. In this case, there exist  generators
$${\cal I}:= (I_1, \dots, I_n) : T^*M \rightarrow B \subset \R^n$$  of
of the Hamiltonian $\R^n$ action so that
each $I_j$ generates a
$2\pi$-periodic Hamiltonian flow.  
The components $I_j$ are called {\it global action variables} and ${\cal I}$ is called a toric moment map.    
In the toric  case, $B$ is a convex polyhedral cone, $B_{reg}$ is simply the interior of $B$, $B_{sing} = \partial B$
(its boundary)  and $m_{cl}(b) \equiv 1.$ Since tori are now labelled by actions, we  write $T_I := {\cal
I}^{-1}(I)$.
Singular orbits $\R^n \cdot (x, \xi)$ are obviously compact non-Lagrangean
tori, and singular levels consist of just one singular orbit.
\medskip

\noindent{\bf Examples}: \\
\noindent(i) $M = \R^n/\Z^n, I_j = \xi_j$, the usual linear coordinates on $T^*(\R^n/\Z^n)$.

\noindent(ii) $M = {\Bbb S}^2, I_1 = p_{\theta}, I_2 = |\xi|_0$, where $p_{\theta}(x, \xi) = \xi(
\frac{\partial}{\partial \theta})$ (the infinitesimal generator of rotations around the
$z$-axis), and where $|\xi|_0$ is the length function of the standard metric.

 \subsection{Riemannian manifolds with completely integrable geodesic flow}

Now suppose that $g$ is a Riemannian metric on $M$ and let $H(x, \xi) =
|\xi|_g$  denote
the associated length function on covectors.   The Hamilton flow $G_t$ of $H$ on
$T^*M-0$ is
homogeneous of degree 1 with respect to the natural $\R^+$ action, and will be referred to as the geodesic flow. It leaves invariant the
 cosphere bundles $S^*M_E = \{H = E\}$ and the flows $G_t^E$ on $S^*M_E$ are all equivalent under
dilation $(x, \xi) \to E (x, \xi)$ to $G^t_1$.

The geodesic flow $G_t$ will be called {\it integrable} if it commutes with a homogeneous Hamiltonian
action of $\R^n$. We may then put $H = p_1$. It is called {\it toric integrable} if it commutes with
 a homogeneous Hamiltonian action of  $\R^n/\Z^n$.  Because $m_{cl}(b) \equiv 1$ in this case, there
 exists a homogeneous function $K$ on $B$ such that $H
= K({\cal I}).$

\noindent{\bf Examples}: The following is a short list of examples:\\

\noindent(i) $M = \R^n/\Z^n$ and $g$ is flat. Then $(M,g)$ is toric integrable.

\noindent(ii) $M = {\Bbb S}^2$ and $g$ is a rotationally invariant metric. If $g$ is of 'simple

type' (e.g. convex), then $(M, g)$ is toric integrable \cite{CV1}.

\noindent(iii) $M = {\Bbb S}^2$ and $g$ is the metric for which  $({\Bbb S}^2, g)$ is an ellipsoid.

\noindent(iv) $M = \R^2/\Z^2$ and $g$ is a Liouville metric (cf. \cite{B.K.S, KMS}).

\noindent (v) Bi-invariant metrics on compact Lie groups. Geodesic flow on $SO(3)$

is known as the Euler top.

\medskip

\subsection{Manifolds without conjugate points}

A Riemannian manifold $(M, g)$ is said to be without conjugate points if there exists a unique geodesic
between each two points of its universal Riemannian cover $(\tilde{M}, \tilde{g})$, or equivalently if every
exponential map $\exp_x : T_x M \to M$ is non-singular.
We will need the following geometric theorems on manifolds without conjugate
points.

\begin{theo} \label{MANE} \cite{M} Let $(M, g)$ be a compact Riemannian
manifold with (co)-geodesic flow
$G^t: T^*M-0 \to T^*M - 0$. Suppose that $G^t$ preserves a  (non-singular)
Lagrangean foliation $\mathcal L$
of $T^*M-0$, i.e. suppose  that $G^t L = L$ for all leaves $L$ of $\mathcal
L$. Then $(M, g)$ has no
conjugate points. \end{theo}

The Hopf conjecture on tori without conjugate points was proved by
Burago-Ivanov:

\begin{theo}\cite{BI}\label{HOPF} Suppose that $g$ is a metric on the
n-torus $T^n$ without conjugate points.  Then $g$
is flat. \end{theo}

\subsection{Integrable Newtonian flows on cotangent bundles}

We will also consider Newtonian flows, i.e. flows of classical Hamiltonians
$H(x, \xi) = \frac{1}{2} |\xi|^2 + V(x)$ on cotangent bundles $T^*M$. Such Hamiltonians and their flows $G_t$ are no
longer homogeneous. The invariant energy surfaces $X_E = \{H = E\}$  and the restricted
flows $G_t^E$ of $G_t$ to $X_E$ may change drastically with $E$. In particular, it may be completely integrable
for some values of $E$ and not others.

\medskip

\noindent{\bf Examples}:\\

\noindent(i) The  spherical pendulum: $M = {\Bbb S}^2$, $H= |\xi|^2 + \cos \phi$; $|\xi|^2$ corresponds

to the round metric and $\phi$ is
the azimuthal angle. 

\noindent(ii) The C. Neumann oscillator on $T^*{\Bbb S}^{n}.    H= |\xi|^2 + \sum_{j=1}^{n} \alpha_{j} x_{j}^{2}$ on $T^* {\Bbb S}^{n}$. Here $0 < \alpha_{1} < ... < \alpha_{n}$ are constants, $(x_{1},...,x_{n})$ are Cartesian coordinates on ${\Bbb R}^{n+1}$ and $|\xi|^2$ corresponds to the usual round metric.

\noindent(iii) The  Kowalevsky and Chaplygin tops \cite{He}. 

\medskip

We note that in the non-homogeneous case, the joint flows $\Phi_t^E$ on each energy level are distinct systems,
and may be integrable for only some values of $E$.  An interesting case is the Chaplygin top \cite{He},  which is
integrable only when the angular momentum integral is put equal to zero.

\subsection{Rigidity theorems for Newtonian flows}

We will need a generalization of Mane's rigidity theorem  to Newtonian flows on tori. The following combines some
ideas of Bialy-Polterovich \cite{BP} and Knauf \cite{K} to give a rigidity result when $M$ is a torus and $H$
is completely integrable with only compact regular orbits. In fact, it is more general:

\begin{prop}\label{BP} Suppose that $g$ is a metric and $V(x)$  is a potential on the
$n$-torus ${\Bbb T}^{n}$ such  that the Hamiltonian flow $G_t^E$  of $H(x,\xi)$ on $X_E$
preserves a $C^{1}$ Lagrangean foliation by tori which project regularly to ${\Bbb T}^{n}$. Then $E > \max V$ and $(E - V) g$ is a flat metric. 

\end{prop}

\begin{proof}  By ( \cite{K}, Theorem 2) no such invariant foliation exists unless $E > \max V$, so we may
assume this is the case. The Jacobi metric $(E - V) g$ is then a well-defined metric on ${\Bbb T}^{n}$. We
denote by $|\xi|^2_{J, E}$ the associated homogeneous Hamiltonian (length squared of a covector).  Since
the sets $\{H = E\}$ and $\{|\xi|^2_{J, E} = 1\}$ are the same, the latter carries a Lagrangean foliation by tori
which project regularly to ${\Bbb T}^{n}.$ Since the geodesic flow $G^t_{J, E}$
of $(E - V)g$  on $\{|\xi|^2_{J, E} = 1\}$ coincides (up to a time  re-parametrization) with  $G_t^E,$   this foliation is invariant under  $G^t_{J, E}$.

Now let $D_r: T^*M- 0 \to T^*M - 0$ be the dilation $D_r (x, \xi) = (x, r \xi).$ Then $D_r: \{|\xi|^2_{J, E} = 1\}
\to \{|\xi|^2_{J, E} = r^2\}$ intertwines the geodesic flows on these sphere bundles (up to constant time reparametrization).   Since $D_r$ is conformally symplectic it also carries the invariant Lagrangean torus foliation of
$\{|\xi|^2_{J, E} = 1\}$ to an invariant Lagrangean torus foliation of $\{|\xi|^2_{J, E} = r^2\}$. It follows
that $T^*M - 0$ carries a Lagrangean torus foliation invariant under the geodesic flow of the Jacobi metric.
By Mane's theorem, the geodesic flow has no conjugate points  and so  by Burago-Ivanov's theorem,
 (E - V)g
must be flat. \end{proof}

\begin{cor}\label{BPcor} With the same notation as above, suppose that there exists an
 interval $[E_0 - \epsilon, E_0 + \epsilon]$ such that, for all  $E \in [E_0 - \epsilon, E_0 + \epsilon]$,  $G_t^E$  
 preserves a  Lagrangean foliation of  by tori which project regularly to ${\Bbb T}^{n}$.
Then: g is flat and $V$ is constant.
\end{cor}

\begin{proof} The assumption implies that $(E - V) g$ is flat for all $E$ in the interval.
Let $R_E$ denote the curvature tensor of $(E - V)g.$ It is clearly a real analytic function
of $E$. Since $R_E \equiv 0$ in $[E_0 - \epsilon, E_0 + \epsilon]$, it must vanish identically.
Therefore the Newton's flow $\Phi_t$ on $T^*T^{n}$ has no conjugate points. By Remark
1.C and Theorem 1.B of \cite{BP}, it follows that $g$ is flat and $V$ is constant.
\end{proof}

\subsection{Semiclassical quantum integrable systems: semiclassical calculus}

We now provide the necessary background  on quantum integrable systems.
 Since we wish to include  quantizations of possibly inhomogeneous Hamiltonians, the proper framework is
that of semiclassical pseudodifferential operators.

First, we introduce symbols. On a  given an open $U \subset {\Bbb R}^{n}$, we say that
$a(x,\xi;\hbar) \in C^{\infty}(U \times {\Bbb R}^{n})$ is  in the symbol
class  $S^{m,k}(U \times {\Bbb R}^{n})$, provided
$$ |\partial_{x}^{\alpha} \partial_{\xi}^{\beta} a(x,\xi;\hbar)| \leq
C_{\alpha \beta} \hbar^{-m} (1+|\xi|)^{k-|\beta|}.$$
\noindent We say that $ a \in S^{m,k}_{cl}(U 
\times {\Bbb R}^{n})$ provided there exists an asymptotic expansion:
$$ a(x,\xi;\hbar) \sim \hbar^{-m} \sum_{j=0}^{\infty} a_{j}(x,\xi)
\hbar^{j},$$
\noindent with  $a_{j}(x,\xi) \in S^{0,k-j}(U \times {\Bbb R}^{n})$. The
 associated $\hbar$-quantization by $Op_{\hbar}(a)$ is
defined locally by the standard formula:
$$Op_{\hbar}(a)(x,y) = (2\pi \hbar)^{-n} \int_{{\Bbb R}^{n}}
e^{i(x-y)\xi/\hbar} \,a(x,\xi;\hbar) \,d\xi.$$
By using a partition of unity, one constructs a corresponding class,
$Op_{\hbar}(S^{m,k})$, of properly-supported $\hbar$-pseudodifferential
operators acting globally on $C^{\infty}(M)$; as is well known, it is
independent of the  choice of partition of unity. Given $a \in S^{m_{1},k_{1}}$ and $b \in
S^{m_{2},k_{2}}$, the composition is given by  $Op_{\hbar}(a) \circ Op_{\hbar}(b) =
Op_{\hbar}(c) + {\cal O}(\hbar^{\infty})$ in $L^{2}(M)$ where locally,
$$ c(x,\xi;\hbar) \sim \hbar^{-(m_{1}+m_{2})} \sum_{|\alpha|=0}^{\infty}
\frac{{(-i\hbar)}^{|\alpha|} }{\alpha !} (\partial_{\xi}^{\alpha} a)\cdot (
\partial_{x}^{\alpha} b).$$

\begin{defn} \label{SCQCI}
We say that the operators $P_{j}^{\hbar} \in Op_{\hbar}(S^{m,k}_{cl}); \,\,j=1,...,n$, generate a semiclassical quantum completely integrable system on $M$ if for each $\hbar$,
$$ \bullet  \,\, \sum_{j=1}^{n} P_{j}^{\hbar *} P^{\hbar}_{j} \,\,\mbox{is jointly elliptic on } \,  T^{*}M, $$ 
$$  \bullet \,\, [P^{\hbar}_{i}, P^{\hbar}_{j}]=0; \,\,\, \forall{1 \leq i,j \leq n},$$
\noindent and the respective semiclassical principal symbols $p_{1},...,p_{n}$ generate a classical integrable system
on $T^{*}M$ with
$dp_{1} \wedge dp_{2} \wedge \cdot \cdot \cdot \wedge dp_{n} \neq 0$ on a dense open subset of $T^{*}M$. We also assume that the finiteness condition (\ref{HYP}) is satisfied. 

\end{defn}

\subsubsection{Examples}

The basic examples we have in mind are where $P_1^{\hbar} = \hbar^2 \Delta + V \in Op_{\hbar}(S^{0,2}_{cl})$ is a Schroedinger operator over
a compact manifold $M$. Examples include:
\begin{itemize}

\item Quantum integrable Laplacians $\Delta$ such as  Laplacians of Liouville metrics on the sphere or torus
\cite{B.K.S} \cite{KMS}, or
of the  ellipsoid \cite{T3}. 

\item Toric integrable Laplacians such  as the  flat Laplacian on ${\bf T}^n$, or Laplacians for surfaces of revolution of `simple type' (see below and
\cite{CV1}).

\item The quantum spherical pendulum $\hbar^2 \Delta + \cos \phi$: $M = S^2$, $\Delta$ is the standard Laplacian, $V = \cos \phi$ where $\phi$ is
the azimuthal angle.  The commuting operator is $\hbar \frac{\partial}{\partial \theta}$, the generator of rotations
around the $z$-axis. 

\item The C. Neumann oscillator on ${\Bbb S}^{n}$. Here the quantum Hamiltonian is the Schroedinger operator $\hbar^{2} \Delta + \sum_{j=1}^{n} \alpha_{j} x_{j}^{2}$ acting on $C^{\infty}({\Bbb S}^{n})$. Here, $\Delta$ is the spherical, constant curvature Laplacian and the potential is the one described above. For the quantized C. Neumann system, one can construct quantum integrals that are all  second-order, real-analytic, semiclassical partial differential operators on the sphere \cite{T3}.

\item The quantized Euler, Lagrange and Kowalevsky tops. The Euler and Lagrange cases are classical \cite{He}, while the quantum Kowalevsky top was shown to be QCI recently by Heckman \cite{He}. Here, the integrals are semiclassical differential operators in the enveloping algebra of $so(3) \triangleright {\Bbb R}^{3}$ defined as follows: Let $E_{1},E_{2},E_{3}$ be the standard Pauli basis of $so(3,{\Bbb R})$ and  $L_{1},L_{2},L_{3}$ be the corresponding left-invariant vector fields defined by:

$$L_{i}(f)(x) := \frac{d}{dt} \{ f(x \, \exp t E_{i} ) \}_{t=0}.$$

\noindent Fix a unit vector $e \in {\Bbb R}^{3}$ and define the $C^{\infty}$ functions on $SO(3)$ by

$$ Q_{i}(x) := \langle x e_{i}, e \rangle.$$

Then, the space of operators generated by $Q_{1},Q_{2},Q_{3},L_{1},L_{2},L_{3}$ can be identified with $so(3) \triangleright {\Bbb R}^{3}$.

\noindent Two of the quantum integrals are the quantized energy Schroedinger operator, $P_{1}:= \frac{1}{4} \hbar^{2} ( L_{1}^{2} + L_{2}^{2} + 2 L_{3}^{2}) - Q_{1}$  and the quantized momentum operator, $P_{2}= \hbar \sum_{j=1}^{3} Q_{j} L_{j}$. In analogy with the classical case, the third quantum integral is a fourth-order  partial differential operator defined as follows: Put $K:= \hbar^{2} ( L_{1} + i L_{2})^{2} + 4 ( Q_{1} + i Q_{2})$. Then, in terms of $K$, $ P_{3} = K K^{*} + K^{*} K - 8 \hbar^{4} ( L_{1}^{2} + L_{2}^{2}).$

\end{itemize}

\vspace{.1in}

Homogeneous quantum completely integrable systems are the special case where $\hbar$ occurs with the same power
in each term and where the usual homogeneous  symbols of the operators are all of order one, e.g. ns $\hbar  \sqrt{\Delta}$ or $\hbar \sqrt{\Delta + V}$. In this case, one could remove $\hbar$
and use the homogeneous symbolic calculus. However, it is often
more convenient  to convert homogeneous systems $P_1, \dots, P_n$ into semiclassical ones by introducing a semiclassical parameter $\hbar$ (with
values in
some sequence $\{ \hbar_{k}; k=1,2,3,... \}$ with $\hbar_{k} \rightarrow 0$) and  semiclassically
scaling the  $P_j$'s:
\begin{equation} \label{QCI2}
 P_{j}^{\hbar} := \hbar P_{j}; \,\,\,\,j=1,2,...,n.
\end{equation}
\noindent When $P_{1} =
\sqrt{\Delta},P_{2},...,P_{n}$ are
classical pseudodifferential operators of order one, then
 $P_{j}^{\hbar}:= \hbar P_{j} \in Op(S^{0,1}_{cl})$ generate the semiclassical
quantum integrable system in the sense of Definition \ref{SCQCI}.

\subsection{Quantum torus actions} (see \cite{GS} for many details on this case). 
Classical torus actions can always be quantized and produce the simplest examples of toric quantum integrable
systems. The classical actions $\{I_j\}$ can  be quantized as commuting pseudodifferential operators
 $\hat{I}_1, \dots, \hat{I}_n$ whose joint spectrum
$$ Sp(\hat{I}_1, \dots, \hat{I}_n) = \Lambda \subset  (\Z^n + \nu)
\cap B$$
is a lattice (translated by a Maslov index).   The simplest case is that of the torus, where
$\hat{I}_j = \frac{\partial}{\partial \theta_j}$ (with $\theta_j$ denoting the usual angular coordinates).
The operators $\sqrt{\Delta + 1/4}, \frac{\partial}{\partial \theta}$ on $S^2$ provide another example. Less obviously,
any convex surface of revolution has a toric integrable Laplacian (cf. \cite{CV1}).

Just as the classical multiplicity $m_{cl}(b) \equiv 1$ in the toric case, so also the  multiplicity $m (\lambda)$ of the joint eigenvalues is
1 for
$|\lambda|$ sufficiently large [CV.1].  Hence up to a finite dimensional
subspace, there
is a  unique (up to unit scalars) orthonormal basis of joint eigenfunctions
$$\hat{I}_j \phi_{\lambda} = \lambda_j \phi_{\lambda},\;\;\;\;\lambda =
(\lambda_1, \dots,
\lambda_n) \in \Lambda.$$

\subsection{Joint eigenvalue ladders}

In the next section, we will study the localization of sequences of  eigenfunctions on level sets of the moment map. To obtain sequences which localize on a given level ${\mathcal P}^{-1}(b)$ it is necessary to choose the corresponding joint eigenvalues to tend in an appropriate

sense to $b$.  Roughly speaking, such  joint eigenvalues form an `eigenvalue ladder'.

The term comes from the toric case, where
 the joint spectrum $\Lambda$ of the action operators  is a semi-lattice (i.e.
the set of lattice points  in a cone)   We define  {\it ladders} (or rays) in a direction $\lambda$ by:
\begin{equation}  {\bf N}_{ \lambda} = \{k \lambda + \nu, k = 0, 
1,2,\dots\} \,  \subset \Lambda. \end{equation}

In the case of quantizations of torus and other Hamiltonian compact group actions,  semiclassical limits are essentially the same as  limits along ladders (cf. \cite{GS}\cite{CV1}).

In the $\R^n$ case, there is usually no optimal choice of the generators $P_j$, and their joint
spectrum is quite far from a lattice. We therefore define  a homogeneous {\it ladder}  of
eigenvalues in the direction  $b=(b^{(1)},b^{(2)},...,b^{(n)}) \in {\Bbb R}^{n}$  to be  a sequence satisfying
\begin{equation} \label{QCI1}
\{ \lambda_k:= (\lambda^{(1)}_{k},...,\lambda^{(n)}_{k}) \in Spec(P_{1},...,P_{n}); \,\forall
j=1,..,n, \,  \lim_{k\rightarrow \infty}\frac{\lambda_{k}}{|\lambda_{k}|}  \, =
\, b\},
\end{equation}
\noindent where $|\lambda_{k}|:= \sqrt{ |\lambda^{(1)}_{k}|^{\,2} + ... +
|\lambda^{(n)}_{k}|^{\,2} }$.

Finally, we introduce a notion of {\it semiclassical ladders}: We 
fix  $0 < \delta <1$,  $b=(b^{(1)},b^{(2)},...,b^{(n)}) \in {\Bbb R}^{n}$, and define the set
\begin{equation} \label{LAD}
 {\bf L}_{b; \delta}(\hbar):= \{  b_{j}(\hbar):= (b_{j}^{(1)}(\hbar), b_{j}^{(2)}(\hbar), \, ... , \, b_{j}^{(n)}(\hbar)) \in \mbox{Spec}(P_{1},...,P_{n}); \,\, | b_{j}(\hbar) - b | \leq C \hbar^{1-\delta} \, \}.
\end{equation}
Here, $b_{j}^{(1)}(\hbar) = E_{j}(\hbar)$.
Taking a sequence  $\hbar \to 0$, the joint eigenvalues in ${\bf L}_{b; \delta}(\hbar)$ form a  sequence tending to $b$ which is the analogue of a homogeneous ladder.

\section{Localization on tori }

One of the main inputs in the proof of the Theorem is the localization  of a ladder
 of  joint eigenfunctions
of a quantum completely integrable system in a regular  direction $b \in B_{reg}$  on the level set ${\cal P}^{-1}(b)$ of the  moment map. In this section, we prove the relevant localization results. We first consider  toric systems,  where level sets are regular and connected and eigenfunctions necessarily localize on individual tori. In the general
 $\R^n$ case,  ladders of eigenfunctions localize on the possibly disconnected level set  ${\cal P}^{-1}(b)$, and
it is a complicated problem to determine how the limit eigenfunction  mass (or `charge') is distributed among the components. To
deal with this problem, we
define a  notion of the charge of a component, and prove that every compact component of ${\cal P}^{-1}(b)$ is charged  by some
sequence of eigenfunctions. This  result will play an important role in the  proof of the Theorem.

\subsection{Toric integrable systems}

Let $A \in \Psi^o(M)$ denote any zeroth order   pseudodifferential operator  and $d\mu_{\lambda}$ denote Lebesgue measure on the Lagrangian torus $T_{\lambda}$.  In the toric case we have the following localization theorem:

\begin{prop} \cite{Z1} \label{TORASYMP} For any ladder $\{k \lambda + \nu: k = 0, 1, 2, \dots\} $ of joint eigenvalues, we have: 
$$ (A \phi_{k \lambda}, \phi_{k \lambda})   =
\int_{T_{\lambda}} \sigma_A d\mu_{\lambda} +
O(k^{-1}).$$\end{prop}

We thus have:

\begin{cor} \label{LOC} For any invariant torus $T_{\lambda} \subset S^*M$, there exists
a ladder $\{\phi_{k \lambda}, k = 0, 1, 2, \dots\}$ of eigenfunctions localizing on
$T_{\lambda}.$ \end{cor}

\subsection{ {\bf $\R^n$-integrable systems} }

The proper generalization of the  toric localization result Proposition (\ref{TORASYMP}) to ${\Bbb R}^{n}$ actions  
 says that ladders of joint eigenfunctions localize on level sets of the
moment map rather than on individual tori. This result is more or less a folk theorem in the physics literature (see \cite{E, Be, Be2}), and the rigorous result is in principle  known  to experts. However, we were unable to find the result  in the literature, so we sketch the proof here. It uses some material on  quantum Birkhoff normal forms from \cite{CV2}.

Let
 $b$ be a regular value of the moment map ${\cal P}$,  let 
$$ {\cal P}^{-1}(b) = \Lambda^{(1)}(b) \cup \dots \cup  \Lambda^{(m_{cl})}(b),$$
\noindent where the $\Lambda^{(l)}(b); l=1,...,m$ are $n$-dimensional Lagrangian
tori, and  $d\mu_{\Lambda^{(j)}(b)}$  denote  the normalize Lesbegue measure on the 
torus $\Lambda^{(j)}(b)$. Let  $  b_j(\hbar) \in {\bf L}_{b,\delta}(\hbar)  $ and define
\begin{equation} \label{c_j}
c_{l}(\hbar; b_j(\hbar)):= \langle Op_{\hbar}(\chi_{l}) \phi_{b_j(\hbar)}, \phi_{b_j(\hbar)} \rangle ; \,\,\,\,l=1,...,m_{cl}(b).
\end{equation}
\noindent We recall that  $\chi_{l}$ is  cutoff function which is  equal to 1 in the neighbourhood $\Omega^{(l)}(b)$ of the torus $\Lambda^{(l)}(b)$ and vanishes on $\cup_{k\neq l} \Omega^{(k)}(b)$. 
\begin{prop} \label{LWL} Let $b \in B_{reg}$, and  let $\{ \phi_{b_{j}(\hbar)} \}$ be a sequence of $L^{2}$-normalizeed joint eigenfunctions of $P_{1},...,P_{n}$ with joint eigenvalues in
the ladder  ${\bf L}_{b, \delta}(\hbar)$ of (\ref{LAD}).  Then,   for any $a \in S^{0,-\infty }$,  we have that as $\hbar \to 0$: 
$$\langle Op_{\hbar}(a) \phi_{b_{j}(\hbar)}, \phi_{b_{j}(\hbar)} \rangle = \sum_{l= 1}^m c_l(\hbar; b_{j}(\hbar))
\int_{\Lambda^{(j)}(b)} a \,\, d\mu_{\Lambda^{(j)}(b)} + {\cal O}(\hbar^{1-\delta}). $$
\noindent Here, $d\mu_{\Lambda^{(j)}(b)}$ denotes Lebesgue measure on $\Lambda^{(j)}(b)$.
\end{prop}

\noindent{\bf Proof}: Let ${\cal L}^{(l)}$ be the pullback of the Maslov line bundle over $\Lambda^{(l)}$ to the affine torus given by $I_{1}^{(l)}= \cdot \cdot \cdot = I_{n}^{(l)}=0$  and $\Omega^{(l)}$ be a sufficiently small neighbourhoodof $\Lambda^{(l)}$ on which there exist action-angle variables $(\theta^{(l)}, I^{(l)})$. According to the quantum Birkhoff normal form (QBNF) construction [CV2], for  $l=1,...,k$ and $j=1,...,n$, there exist
$\hbar$-Fourier integral operators, $U^{(l)}_{b,\hbar}: C^{\infty}(M) \rightarrow
C^{\infty}({\Bbb T}^{n}; {\cal L}^{(l)})$, microlocally elliptic
 on $\Omega^{(l)}$, together with  $C^{\infty}$ symbols,
$f_{j}^{(l)}(x;\hbar) \sim \sum_{k=0}^{\infty} f^{(l)}_{jk}(x) \hbar^{k}$,
with $f_{j0}(0)=0$ such that:
\begin{equation} \label{QBNF}
U^{(l)*}_{b,\hbar} f_{j}^{(l)}(P_{1}-b^{(1)},...,P_{n}-b^{(n)};\hbar)U^{(l)}_{b,\hbar} =_{\Omega_{0}^{(l)}}
\frac{\hbar}{i} \frac{\partial}{\partial \theta_{j}}.
\end{equation}
\noindent  Moreover, when $P_{1},...,P_{n}$ are  self-adjoint, the
operator $U^{(l)}_{b}$ can be taken to be microlocally unitary.

We now observe that the space of  admissible \cite{CP} solutions of the microlocal eigenfunction equation
\begin{equation}\label{MICRO}
 P_{k} \phi_{b_{j}(\hbar)} =_{\Omega^{(l)}(b)}  b^{(k)}_{j}(\hbar) \phi_{b_j(\hbar)}
\end{equation}
is one-dimensional. Indeed, such solutions are the same as solutions of
$$f_{k}^{(l)}(P_{1}-b^{(1)},...,P_{n}-b^{(n)};\hbar) \phi_{j} =_{\Omega^{(l)}(b)} f_{k}^{(l)}(b^{(1)}_{j}-b^{(1)},...,b^{(n)}_{j}-b^{(n)};\hbar)  \phi_{j}. $$
We conjugate this equation  to   Birkhoff normal form  (\ref{QBNF})  and use the  fact that  the microlocal solutions of the model equation
$$ \frac{\hbar}{i} \frac{\partial}{\partial \theta} u_{j} = m_{j} \,u_{j}$$
\noindent are just multiples of $\exp [ i ( n + \pi \gamma/4) \theta ]$, where $\gamma$ is the Maslov index and $n \in \Bbb Z$. Thus, the joint eigenfunctions $\phi_{b_{j}(\hbar)}$ are given microlocally by
\begin{equation} \label{CHAR}
\phi_{b_{j}(\hbar)} =_{\Omega^{(l)}(b)} \, \sqrt{c_{l}(\hbar;b_{j}(\hbar))} \,\, U^{(l)}_{b;\hbar} ( e^{i(n_{j}+\pi \gamma/4) \theta} ).
\end{equation}
 The right sides  of (\ref{CHAR}) are the usual quasimodes or semiclassical Lagrangian distributions [CV2]

Now let  $\chi_{l}(x,\xi) \in C^{\infty}_{0}(T^{*}M); l=1,....,m_{cl}(b)$ be a cutoff function which is identically equal to one on the neighbourhood $\Omega^{(l)}(b)$ and vanishes on $\Omega^{(k)}(b)$ for $k \neq l$. For  $\hbar$ sufficiently small, we then have
$$ \langle Op_{\hbar}(a) \phi_{b_{j}(\hbar)}, \phi_{b_{j}(\hbar)} \rangle = \sum_{l=1}^{m_{cl}(b)} \langle Op_{\hbar}(a) \circ Op_{\hbar}(\chi_{l}) \phi_{b_{j}(\hbar)}, \phi_{b_{j}(\hbar)} \rangle + {\cal O}(\hbar^{\infty}).$$
It follows by (\ref{CHAR}), the 
semiclassical Egorov theorem and a Taylor expansion about the Lagrangian  torus $I^{(l)}=0$  that:
\begin{equation}\begin{array}{lll} \label{sec452}
 \langle Op_{\hbar}(a) \circ Op_{\hbar}(\chi_{l}) \phi_{b_{j}(\hbar)},  \phi_{b_{j}(\hbar)} \rangle &
= & c_l(\hbar;b_{j}(\hbar))
 \langle Op_{\hbar}(a) \circ Op_{\hbar}(\chi_{l}) U^{(l)}_{b;\hbar} ( e^{i(n_{j}+\pi \gamma/4) \theta} ) ,
 U^{(l)}_{b;\hbar} ( e^{i(n_{j}+\pi \gamma/4) \theta} ) \rangle \\ & & \\
& = &  c_l(\hbar;b_{j}(\hbar)) \langle U^{(l) * }_{b;\hbar} Op_{\hbar}(a) \circ Op_{\hbar}(\chi_{l}) U^{(l)}_{b;\hbar}  e^{i(n_{j}+\pi \gamma/4) \theta}  ,    e^{i(n_{j}+\pi \gamma/4) \theta}  \rangle \\ &  & \\ &= & 
(2\pi)^{-n} \, c_{l}(\hbar;b_{j}(\hbar)) \, \left( 
\int_{\Lambda^{(l)}} a \, \,d\mu_{l}  + e(\hbar) \right) + {\cal O}(\hbar),
\end{array} \end{equation}
 where 
 $ e(\hbar) = \langle  Op_{\hbar}(r) u_{\hbar}, u_{\hbar} \rangle$ for some function  $r \in C^{\infty}_{0} ({\Bbb T}^{n} \times D_{1})$ satisfying
$r(\theta,I) = {\cal O}(|I|)$ (recall, we have normalized the action variables so that $I^{(l)}=0$ on the torus $\Lambda^{(l)}(b)$).  Here, $u_{\hbar}(\theta) = \exp [i(m_{1} \theta_{1} + ...+ m_{n}\theta_{n})]$ with $m_{j}(\hbar) = {\cal O}(\hbar^{1-\delta})$.

An integration by parts in  the $I_{1},...,I_{n}$
variables  shows that:
$$( Op_{\hbar}(r) u_{\hbar}, u_{\hbar} ) = {\cal O}(\hbar^{1-\delta}),$$

\noindent and the proposition follows.\qed

\subsection{Charge of compact Lagrangean  orbits}

We now investigate the coefficients  $c_j(\hbar)$ in Proposition (\ref{LWL}) for `ladders' of eigenfunctions.
Our purpose is to show that there exist ladders for which the limit as $\hbar \to 0$ of $c_j(\hbar)$ is bounded
below by a positive geometric constant.  It is convenient at this point to introduce the language of quantum limits.

\subsubsection{Quantum limits}  Let $(P_1, \dots, P_n)$ denote a quantum integrable system, with
classical integrable flow $\Phi_t$. Fix $E$ and let  $M_I^E$ denote the set of invariant probability measures for $\Phi_t^E$
on $ X_{E}$.  For instance, $M_I^E$ includes the
 orbital averaging
measures $\mu_z$, defined by $$\int_{X_{E}} f d \mu_z = \lim_{T \to \infty}
\frac{1}{T^n} \int_{\max |t_j| \leq T}  \linebreak f(\Phi_t(z)) dt.$$ In the case of compact
(torus) orbits, $\mu_z$ is  the Lebesgue probability measure on the orbit of $z$.

By the set ${\mathcal Q}_E $  of `quantum limit' measures of the quantum integrable
system at energy level $E$, we mean
the set of weak* limits (as $\hbar \to 0$) of the measures  $d \Phi_{b_{j}(\hbar)}$ defined by
\begin{equation} \langle Op_{\hbar}(a) \phi_{b_{j}(\hbar)}, \phi_{b_{j}(\hbar)}\rangle =
\int_{X_E} a \, \, d \Phi_{b_{j}(\hbar)}, \;\;(b_j^{(1)} \to E). \end{equation}
We write $d \Phi_{b_j(\hbar)} \to d\mu \in {\mathcal Q}_E$ for weak* convergence to the limit  as $\hbar \rightarrow 0$.
It is an
easy consequence of the semiclassical Egorov theorem that ${\mathcal Q}_E \subset M_I^E.$
 When $d\mu$ equals
Lebesgue probability measure on an orbit, we say that the sequence $\{\phi_{b_j(\hbar)}\}$ localizes on the orbit.
For background,
 terminology and references in a closely related context,   we refer
to \cite{JZ}.

We now consider quantum limits of eigenfunctions corresponding to a ladder of joint eigenvalues. Put:
\begin{equation} \label{EFBOX} {\bf V}_{b, \delta}(\hbar) = \{\phi_{b_j(\hbar)}:   b_{j}(\hbar) \in  {\bf L}_{b; \delta}(\hbar)  \} \end{equation}
There are many possible weak* limit points of the set $\cup_{\hbar \in [0, \hbar_0]} {\bf V}_{b, \delta}(\hbar).$
We say:
\begin{defn}\label{EFNLAD}  For $b \in B_{reg}$, a ladder of eigenfunctions is a sequence  ${\mathcal E}_b:= \{\phi_{b_j}(\hbar) \}$  of joint eigenfunctions with the following properties:
\begin{itemize}

\item  $b_j (\hbar) \in  {\bf L}_{b,\delta}(\hbar) $ as $\hbar \to 0$  forms an eigenvalue ladder;

\item  $d\Phi_{b_j}(\hbar) $ has a unique weak limit $d \Phi_{{\mathcal E}_b}$ as $\hbar \to 0.$
\end{itemize}\end{defn}

For a ladder of eigenfunctions, $\lim_{\hbar \to 0} c_{\ell}(\hbar;b_{j}(\hbar))$ exists for each $\ell$ in Proposition (\ref{LWL}).

\begin{defn} \label{CHARGE} Given  $b \in B_{reg}$, we  say that the ladder ${\mathcal E}_b  = \{ \phi_{b_{j}(\hbar)} \}$
 gives  {\it charge}  $c_l({\mathcal E}_{b}) := \lim_{\hbar \to 0} c_{\ell}(\hbar;b_{j}(\hbar))$ to the  component torus $\Lambda^{(l)}(b)$, and that it {\it charges}   $\Lambda^{(l)}(b)$
if $c_l({\mathcal E}_{b})  > 0.$ \end{defn}

The limit in Definition (\ref{CHARGE}) above clearly depends on the ladder  ${\mathcal E}_b$. For instance, there could be sequences of joint eigenfunctions localizing on each
single component of ${\mathcal P}^{-1}(b).$ To obtain an invariant of the Lagrangean orbits which is independent
of the ladder, we say:

\begin{defn} \label{charge}  The {\it charge} $c(\Lambda^{(l)}(b))$ of a component torus $\Lambda^{(l)}(b) \subset {\mathcal P}^{-1}(b)$ is defined by 
by the formula:
$$\begin{array}{lll}    c (\Lambda^{(l)}(b)) &=& \sup_{{\mathcal E}_b}  c_l({\mathcal E}_b)  \end{array}$$
where $c_l$ is the coefficient in the sum of Proposition (\ref{LWL}). \end{defn}

A useful formula for the charge is:

\begin{prop} \label{equivch}
$  c(\Lambda^{(l)}(b))   = \limsup_{\hbar \rightarrow 0}  \max_{ \phi_{b_j(\hbar)} \in  V_{\delta}(\hbar)}  \langle Op_{\hbar}(\chi_{l}) \phi_{b_{j}(\hbar)}, \phi_{b_{j}(\hbar)} \rangle.  $\end{prop}

\begin{proof}

\noindent(i) $\geq$: By definition,  $  c_l(\hbar; b_{j}(\hbar))  = \langle Op_{\hbar}(\chi_{l}) \phi_{b_{j}(\hbar)}, \phi_{b_{j}(\hbar)} \rangle$ where
 $\chi_{l}$ is a cutoff to $\Omega_{l}.$ Since $V_{b, \delta}(\hbar)$ is a finite set for each $\hbar$, there exists
 $\phi_{b_j(\hbar)}^{\max} \in V_{b, \delta}(\hbar)$
such that $    \langle  Op_{\hbar}(\chi_{l}) \phi^{\max}_{b_{j}(\hbar)}, \phi^{\max}_{b_{j}(\hbar)} \rangle   =
  \max_{ \phi_{b_j(\hbar)} \in  V_{\delta}(\hbar)}  \langle Op_{\hbar}(\chi_{l}) \phi_{b_{j}(\hbar)}, \phi_{b_{j}(\hbar)} \rangle.$
We form the sequence $\{\phi_{b_j(\hbar)}^{\max}\}_{\hbar \in \{ \hbar_{k} \} }$ and then choose a sub-ladder ${\mathcal E}_b^{\max}$ with a unique quantum limit.
Then
$$\begin{array}{l}  c(\Lambda^{(l)}(b))  \geq  \lim_{\hbar \rightarrow 0}  \langle Op_{\hbar}(\chi_{l}) \phi_{b_{j}(\hbar)}^{max}, \phi_{b_{j}(\hbar)}^{max} \rangle \\ \\
\geq \limsup_{\hbar \rightarrow 0}  \max_{ \phi_{b_j(\hbar)} \in  V_{\delta}(\hbar)}  \langle Op_{\hbar}(\chi_{l}) \phi_{b_{j}(\hbar)}, \phi_{b_{j}(\hbar)} \rangle. \end{array}$$

\noindent(ii) $\leq$: It is clear that for each ladder ${\mathcal E}_b$ we have
$$  c_l({\mathcal E}_b)   \leq  \limsup_{\hbar \rightarrow 0}  \max_{ \phi_{b_j(\hbar)} \in  V_{\delta}(\hbar)}  \langle Op_{\hbar}(\chi_{l}) \phi_{b_{j}(\hbar)}, \phi_{b_{j}(\hbar)} \rangle.$$
Therefore the same holds after taking the supremum over ${\mathcal E}_b.$

\end{proof}

The following lemma is the main result of this section:

\bigskip

\begin{lem} \label{TC}  Let $\omega^{(l)}(b)$ denote Liouville measure of the Lagrangian torus $\Lambda^{(l)}(b); \, l=1,...,m_{cl}(b)$. Then,  for all $(b,l) \in B_{reg} \times \{1,...,m_{cl}(b) \}$ we have that  
$$c(\Lambda^{(l)}(b)) \geq \frac{ \omega^{(l)}(b)} { \sum_{j=1}^{m_{cl}(b)} \omega^{(j)}(b) }. $$
\end{lem}
\noindent{\bf Proof:}  Fix $\zeta \in {\cal S}({\Bbb R}^{n})$ with $\zeta \geq 0$, $\check{\zeta} \in C^{\infty}_{0}({\Bbb R}^{n})$ and $\check{\zeta}(0)=1$.  Assume moreover that $0 \in {\Bbb R}^{n}$ is the only point of intersection of supp $\zeta$ with the joint periods of the joint flow $\Phi_{t}$. 
 Let $K$ be a fixed compact neighbourhood of $b=(b^{(1)},...,b^{(n)})$ and $a \in S^{0,-\infty}$. Consider the localized semiclassical trace:
\begin{equation}\label{TC1}
Tr_{a} (\zeta):= \sum_{ b_{j}(\hbar) \in K} \langle Op_{\hbar}(a) \phi_{b_{j}(\hbar)}, \phi_{b_{j}(\hbar)} \rangle \,\, \zeta \, \left( \frac{b_{j}(\hbar) - b}{\hbar} \right).
\end{equation}

\noindent The localized semiclassical trace formula for commuting operators [Ch] implies that for any $a \in S^{0,-\infty}$ and $\zeta \in {\cal S}({\Bbb R}^{n})$ as above,
\begin{equation}\label{TC1.5}
 Tr_{a}(\zeta) = (2\pi)^{-n} \, \int_{ {\cal P}^{-1}(b)} \, a \, d\omega^{(l)}(b)   + {\cal O}(\hbar).
\end{equation}

\noindent So, in particular putting $a(x,\xi)  = \chi_{l} (x,\xi)$, we have that:
\begin{equation} \label{TC2}
Tr_{\chi_{l}} (\zeta) = (2\pi)^{-n}  \, \int_{\Lambda^{(l)}(b)} \chi_{l} \, d\omega^{(l)}(b)   + {\cal O}(\hbar) = (2\pi)^{-n} \,  \omega^{(l)}(b)  + {\cal O}(\hbar),
\end{equation}
\noindent  since $\chi_{l}=1$ on the torus, $\Lambda^{(l)}(b)$. On the other hand, since $\zeta \in {\cal S}({\Bbb R}^{n})$, it follows that
\begin{equation}\label{trun} 
Tr_{\chi_{l}} (\zeta)= \sum_{ \{ b_{j}(\hbar) \in {\bf L}_{b,\delta}(\hbar) \} }  \langle Op_{\hbar}(\chi_{l}) \phi_{b_{j}(\hbar)}, \phi_{b_{j}(\hbar)} \rangle \, \zeta \, \left( \frac{b_{j}(\hbar)- b}{\hbar} \right) \, + \, {\cal O}(\hbar^{\infty}).
\end{equation}
\noindent Thus, by the definition (\ref{charge}) of the charge $c(\Lambda^{(l)}(b)$ and the fact that $\zeta \geq 0$, we have that:
\begin{equation} \label{TC5}
 | Tr_{\chi_{l}} (\zeta) |  \leq  (2\pi)^{-n} \, \left( \max_{  \{   b_{j}(\hbar) \in {\bf L}_{b,\delta}(\hbar)   \} } \langle Op_{\hbar}(\chi_{l}) \phi_{b_{j}(\hbar)} , \phi_{b_{j}(\hbar)} \rangle \right)  \,  \sum_{ \{ b_{j}(\hbar) \in {\bf L}_{b,\delta}(\hbar)\} }   \zeta \, \left( \frac{b_{j}(\hbar)- b}{\hbar} \right) \, + \, {\cal O}(\hbar^{\infty}). 
\end{equation}

\noindent Next, by applying the trace formula once again we get that:
\begin{equation} \label{TC6}
\sum_{ \{ b_{j}(\hbar) \in {\bf L}_{b,\delta}(\hbar) \} }   \zeta \, \left( \frac{b_{j}(\hbar)- b}{\hbar} \right) \, = (2\pi)^{-n} \, \sum_{j=1}^{m_{cl}(b)} \omega^{(j)}(b) \, + {\cal O}(\hbar).
\end{equation}
\noindent Substituting (\ref{TC6}) in (\ref{TC5}) yields the estimate
\begin{equation} \label{TC7}
| Tr_{\chi_{l}} (\zeta) |  \leq  (2\pi)^{-n} \,  \max_{  \{ b_{j}(\hbar)  \in {\bf L}_{b,\delta}(\hbar) \} } \langle Op_{\hbar}(\chi_{l}) \phi_{b_{j}(\hbar)} , \phi_{b_{j}(\hbar)} \rangle   \cdot \left( \sum_{j=1}^{m_{cl}(b)} \omega^{(j)}(b) \right) \,\,  + {\cal O}(\hbar).
\end{equation}
\noindent The lemma then follows by combining (\ref{TC7}) and (\ref{TC2}) and letting $\hbar \rightarrow 0$.
\qed

This yields a generalization of Corollary (\ref{LOC}):
\begin{cor}\label{ch}  For any $b \in B_{reg}$, and 
for any   $1 \leq \ell \leq m_{cl}(b)$, 
there exists a ladder  ${\mathcal E}_{b}^{(\ell)} = \{ \phi_{b_{j}(\hbar)} \}    $ such that $  c_{\ell}({\mathcal E}_{b}^{(\ell)}) \geq  \frac{ \omega^{(\ell)}(b)} { \sum_{j = 1}^{m_{cl}(b)}
\omega^{(j)}(b) } ;$

 \end{cor}
Thus,  every regular torus orbit is charged by some ladder.  This follows from Lemma ( \ref{TC}), Proposition ( \ref{equivch}) and Proposition ( \ref{LWL}).

\subsubsection {Charge of compact singular orbits}

Our next step is to prove that some compact singular orbits are also charged. To be precise, we have so far only defined the notion of charge for regular levels of the moment map (Definition (\ref{CHARGE}). The analogous defintion in the case of a singular value $b_s \in B_s$  is as follows. Let ${\cal P}^{-1}(b_{s}) = \cup_{j = 1}^r  \Gamma_{sing}^{(j)}(b_s)$ be the decomposition (\ref{SL}) into connected components.

\begin{defn} \label{SINGCHARGE}  When $b_{s} \in B_{sing}$, we define an eigenfunction ladder ${\mathcal E}_{b_s}$ to be a sequence of joint eigenfunctions with joint eigenvalues satisfying  $b_{j}(\hbar) - b_{s}= o(1)$ as $\hbar \rightarrow 0$  and with unique limit measure  $d \Phi_{{\mathcal E}_{b_s}}$. We say that ${\mathcal E}_{b_s}$ gives charge $\int_{\Gamma_{sing}^{(j)}(b_s)} d\Phi_{{\mathcal E}_{b_s}}$ to the component $\Gamma_{sing}^{(j)}$. Similarly, we say that it gives charge $\int_{\Lambda_{sing}^{(j)}(b_s)} d\Phi_{{\mathcal E}_{b_s}}$ to any orbit $\Lambda_{sing}^{(j)}(b_s)$  on $\Gamma_{sing}^{(j)}(b_s)$ (see (\ref{ORB})).  Finally,  the {\it charge} $c(\Gamma^{(j)}(b_s)),$ resp. $c(\Lambda^{(j)}(b_s))$ of a component, resp.   an orbit on the component, is the supremum of the same over all ladders $ {\mathcal E}_{b_{s}} $\end{defn}

We then have: 
\begin{lem} \label{CS1}  Let $b_s \in B_{sing}$,  and let  $\{ \Gamma_{sing}^{(j)}(b)\}$ denote the singular  components of ${\cal P}^{-1}(b_{s})$. Then, there exists $j$ such that
$c(\Gamma_{sing}^{(j)}(b)) > 0$. 
 Further, there exists a compact singular orbit $\Lambda^{(j)}(b_s) \subset \Gamma_{sing}^{(j)}(b)$ such that $c(\Lambda^{(j)}(b_s)) > 0.$ \end{lem}

\begin{proof}
Let $U_{sing}$ be a $\Phi_t$-invariant neighbourhood of $\cup_{j = 1}^r \Gamma_{sing}^{(j)}(b) $. Let  $\{b_n\} \subset  B_{reg}$ be a sequence of regular points such that $b_n \to b_{s}$.  For each $j$ and sufficiently large $n$, there  exists at least one component $\Lambda^{(\ell)}(b_n)$ of ${\cal P}^{-1}(b_n)$ such that   $ \Lambda^{(\ell)}(b_n) \subset U_{sing}$. By Lemma  (\ref{TC}), 
 $\Lambda^{(\ell)}(b_n)$ is charged by an amount $\geq \frac{\omega^{(\ell)}(b_n)}{\sum_{j = 1}^{m_{cl}(b_n)} \omega^{(j)}(b_n) }$.  

We now break up the   discussion  into two cases:
\bigskip

\noindent{\bf Case 1: All $\R^n$-orbits of $\cup_{j = 1}^r \Gamma_{sing}^{(j)}(b) $ are compact}

In this case, we just need a positive lower bound for the quotient $\frac{\omega^{(\ell)}(b_n)}{\sum_{j = 1}^{m_{cl}(b_n)} \omega^{(j)}(b_n) }$ as $n \to \infty.$  A lower bound for the numerator is given by the minimal period of
Yorke's theorem (Proposition \ref{YORKE}). Since all orbits (including the limit) are compact, the masses in the
denominator have uniform upper bounds. Indeed, by (\ref{LM}) the masses are the co-volumes of the period lattices
of $\Lambda^{(\ell)}(b).$ Since the period vectors generating the lattices are uniformly bounded as $n \to \infty$,
the volumes are also uniformly bounded above.  Hence the denominator is bounded above, and therefore the quotient
is bounded below by a positive constant. \qed

\bigskip

\noindent{\bf Case 2: There exists a non-compact orbit in $\cup_{j = 1}^r \Gamma_{sing}^{(j)}(b) $}

In this case, the denominator will tend to infinity, so we need a better lower bound on the numerator. 
We claim  that there exists $\ell$ such that 
 $ \Lambda^{(\ell)}(b_n) \subset U_{sing}$ and $c(\Lambda^{(\ell)}(b_n) ) \geq \frac{1}{m_{cl}(b_n) }$.
To prove this, it suffices to find  $\ell$ such that 
\begin{equation}\label{MAX} \frac{\omega^{(\ell)}(b_n)}{\sum_{j:  \Lambda^{(j)}(b_n) \subset U_{sing}} \omega^{(j)}(b_n) } 
\geq \frac{1}{\# \{j:  \Lambda^{(j)}(b_n) \subset U_{sing}\} }. \end{equation}
The natural candidate is to choose $\ell$ such that 
\begin{equation} \label{MAXCHOICE} \omega^{(\ell)}(b_n) = \max_{\{j:  \Lambda^{(j)}(b_n) \subset U_{sing}\}} \omega^{(j)}(b_n). \end{equation} 
We now prove that this choice of $\ell$ satisfies (\ref{MAX}).

We  write $$\sum_{j = 1}^{m_{cl}(b_n)} \omega^{(j)}(b_n)  = \sum_{j:  \Lambda^{(j)}(b_n) \subset U_{sing}} \omega^{(j)}(b_n) + 
\sum_{j:  \Lambda^{(j)}(b_n) \cap U_{sing} = \emptyset} \omega^{(j)}(b_n).$$
The second term is bounded above by a constant $C$ independent of $n$.  
 The first term
tends to infinity since $\cup_{j = 1}^r \Gamma_{sing}^{(j)}(b) $ contains a non-compact
orbit.  Indeed, at least
one vector of the period lattice of $\Lambda^{(j)}(b_n)$ must tend to infinity as $n \to \infty$ since the limit
orbit is non-compact. It follows that the set of period lattices ${\mathcal I}^{(j)}_b$ is non-compact in the
manifold of lattices of full rank of $\R^n$. Now according to Mahler's theorem, any set 
$$\{\Gamma \subset \R^n |   ||\gamma|| \geq C,\;\; (\gamma \in \Gamma -\{0\}),\;\;\mbox{and}\;\;
  Vol(\R^n/\Gamma) \leq K  \}$$
is compact. 
By Yorke's theorem (loc. cit.), the minimal period stays bounded below, so non-compactness of the lattices
forces some volume $\omega^{(\ell)}(b_n) \to \infty$ as $n \to \infty.$

It follows that when a non-compact orbit exists in ${\mathcal P}^{-1}(b_s)$, then for each $\ell$, 
$$\frac{\omega^{(\ell)}(b_n)}{\sum_{j = 1}^{m_{cl}(b_n)} \omega^{(j)}(b_n) } = 
\frac{\omega^{(\ell)}(b_n)}{\sum_{j:  \Lambda^{(j)}(b_n) \subset U_{sing}} \omega^{(j)}(b_n) }+ 
o(1) \,\,\,\,\,  \mbox{as}\,\, n \rightarrow \infty. $$
Then (\ref{MAX}) follows  if  we select $\ell$ as in (\ref{MAXCHOICE}). \qed
\bigskip

We now complete the proof of Lemma (\ref{CS1}).
 By the finite complexity condition, we have found $ \Lambda^{(j_n)}(b_n) \subset U_{sing}$ such that $c(\Lambda^{(j_n)}(b_n) ) \geq c:= \frac{1}{M} > 0$. Further, for each $n$, there exists a ladder ${\mathcal E}_{b_n}$ which  gives charge $ \geq c$ to  $\Lambda^{(j_n)}(b_n)
\subset {\cal P}^{-1} \cap U_{sing}$. Let $d \Phi_{{\mathcal E}_{b_n}}$ denote the unique weak limit measure of the ladder. Then let $\nu$ denote any weak* limit of the sequence  $\{ d \Phi_{{\mathcal E}_{b_n}}\}$. It follows that  $\nu$ is  an invariant probability measure supported on  $\cup_{j = 1}^{r} \Gamma_{b_s}^{(j)}.$ Indeed, its support must be contained in  the set of  limit points  of  the sequence of orbits $\{\Lambda^{(j_n)}(b_n)\}$, hence in ${\mathcal P}^{-1}(b_s) \cap U_{sing}.$ Since ${\cal Q}$ is closed in the weak* topology (since it is a set of limit points), it follows further  that  $\nu \in {\cal Q}.$ Hence there exists a  ladder ${\mathcal E}_{b_s}$ such that $\Phi_{{\mathcal E}_{b_s}} \to \nu,$ and which charges $\cup_{j = 1}^{r} \Gamma_{b_s}^{(j)}$ by an amount $c > 0.$  This proves the first part of the lemma. The second statement is an immediate consequence of Proposition(\ref{EXISTS}): There must exist at least one  compact singular orbit  $\Lambda_{b_s}^{(j)} \subset \cup_{j = 1}^{r} \Gamma_{b_s}^{(j)}.$ Since $\nu$ is an invariant probabililty measure, it must be supported on union of the compact singular
orbits, hence must charge at least one such orbit.
\end{proof}

\section{ Proof of the Theorem }

We break up the proofs into three steps. Step 1 is to show that the uniform boundedness assumption implies
that all regular tori project without singularities to the base. Step 2 is to show that there are no singular
tori.  Step 3 is a geometric argument showing that any completely integrable system with no singular tori
and with all tori projecting regularly to the base is flat. 

\subsection{Step 1: regular tori project regularly}

We first consider the simplest case of toric systems:

\subsubsection{Toric integrable systems}

\begin{prop} \label{nonsing} Suppose that $(M,g)$ is toric integrable and
that
 $L^{\infty}(E, g) = O(1).$ Then  every
orbit of the torus action  has a non-singular projection to $M$.
In particular, the orbit foliation is a non-singular Lagrangean foliation.
\end{prop}

\noindent{\bf Proof}: The assumption implies that the joint eigenfunctions
$\{\phi_{\lambda}\}$ of
the quantum torus action have uniformly bounded sup-norms.

By Proposition (\ref{TORASYMP}), 
for every invariant torus $T_{\lambda}$, there exists a ladder $\{k \lambda, k = 1, 2, \dots\}$ of joint eigenvalues
such that for all  $V \in C^{\infty}(M)$
we have
$$\lim_{k \rightarrow \infty} \int_{M} V(x) |\phi_{k \lambda}(x)|^2 dvol  =
\int_{M} V \pi_{\lambda *} d\mu_{\lambda}.$$
If we have  $||\phi_{k \lambda}||_{\infty} \leq C$ for all $(k, \lambda)$, then
$$ |\int_{M} V(x) |\phi_{k \lambda}(x)|^2 dvol|  \leq C
||V||_{L^1}\;\;\;\;\; (\forall k)$$
 and hence
$$\lim_{k \rightarrow \infty} |\int_{M} V(x) |\phi_{k \lambda}(x)|^2 dvol|
\leq C ||V||_{L^1}.$$
Therefore
\begin{equation} \label{LONE} |\int_{M} V\pi_{\lambda *}d\mu_{\lambda} | \leq C ||V||_{L^1} \end{equation}
which implies that $\pi_{\lambda *}d \mu_{\lambda}$ is a continuous linear
functional on $L^1(M)$, hence
belongs to $L^{\infty}(M)$.  That is, we may write 
$\pi_{\lambda *}d \mu_{\lambda}  = f_{\lambda} dvol$,  with
 $||f_{\lambda}||_{\infty}
\leq C.$  If $\pi_{\lambda}$ had a singular value, it is easy to check that
$\pi_{\lambda *}d \mu_{\lambda}$ would
blow up there .   Hence, $\pi_{\lambda}$ is a non-singular projection.
\qed
\medskip

Now we turn to the general case:

\subsubsection{$\R^n$ actions}

\begin{prop} \label{TB} All regular tori project diffeomorphically to the base. \end{prop}

\noindent{\bf Proof:} Since by Lemma (\ref{TC}) a regular torus $\Lambda^{(l)}(b)$  has charge $ c(\Lambda^{(l)}(b)) \geq \frac{ \omega^{(l)}(b)} { \sum_{j=1}^{m_{cl}(b)} \omega^{(j)}(b) }>  0 $, it follows by Corollary (\ref{ch}) that there exists a ladder of joint 
eigenfunctions $  \{ \phi_{b_{j}(\hbar)} \} \subset {\mathcal E}_{b} $ with the property that:
$$\langle V \phi_{b_{j}(\hbar)}, \phi_{b_{j}(\hbar)} \rangle =  \sum_{l=1}^{m_{cl}(b)}  c_{l}({\mathcal E}_{b}) \,  \int_{\Lambda^{(j)}(b)} V d\mu_{\Lambda^{(j)}(b)} + o(1),$$
\noindent where  $ c_{l}({\mathcal E}_{b}) \geq \frac{ \omega^{(l)}(b)} { \sum_{j=1}^{m_{cl}(b)} \omega^{(j)}(b) } >0 $ and $c_{k}({\mathcal E}_{b}) \geq 0$ for $k \neq l$. Thus,  we have (as in the toric case) that 
$$c_{l}({\mathcal E}_{b}) \,\,   \int_{M} V \pi_{*}d\mu_{\Lambda^{(j)}(b)} \leq C\, ||V||_{L^1},$$
\noindent where we can take $C= c_{l}({\mathcal E}_{b}) \,  \cdot L^{\infty} (\hbar,b_{j}(\hbar);g,V).$
Since $c_{l}({\mathcal E}_{b}) > 0$ we can cancel it to find that the torus projects regularly. 
\qed

\noindent As an immediate of Proposition (\ref{TB}) we have:

\begin{cor}\label{UB} Let $\{ \pi_* d \mu_{\Lambda}\}$ denote the set of
projections to $M$ of normalized Lebesgue measures on compact Lagrangean
tori $\Lambda \subset X_{E}.$ Then, under the assumptions of Theorem (\ref{RM}), the family is uniformly bounded as linear
functionals on $L^{1}(M).$ \end{cor}

\subsection{Non-existence of singular levels}

 We have:

\begin{lem} \label{NOSL} Under the assumptions of Theorem (\ref{RM}), ${\cal P}$ has
no singular levels; all orbits are Lagrangean.  \end{lem}

\begin{proof}

Existence of a compact singular orbit contradicts the
 the uniform boundedness of eigenfunctions assumption.  Indeed, it follows from  Lemma (\ref{CS1})
that,  for any $V \in C^{\infty}(M)$,  there exist a compact, singular orbit $\Lambda_{sing}^{(l)}$ and $L^{2}$-normalized joint eigenfunctions $\{ \phi_{b_{j}(\hbar)} \}$ such that for some $c ( \Lambda_{sing}^{(l)}  ) > 0,$ 
\begin{equation}\label{pf1}
  c ( \Lambda_{sing}^{(l)}  ) \,  \int_{\Lambda_{sing}^{(j)}} V \, \pi_{*} d\nu_{l} \, \leq C  \| V \|_{L^{1}(M)} .
\end{equation}
\noindent However, the estimate in (\ref{pf1})  cannot hold since by definition, compact singular orbits
have dimension $\dim \Lambda_{sing}^{(l)} <n$. Therefore, there cannot exist singular levels of the moment map ${\cal P}$.
\end{proof}

\subsection{Completion of proof of Theorem}

We first complete the proof of  Theorem (\ref{RM}) for general metrics with quantum  completely integrable Laplacians. 
 Subsequently
we take up the case of Schroedinger operators.

The first step is to consider projections of regular Lagrangean tori. By Proposition(\ref{TB}), the assumption
of uniformly bounded eigenfunctions then applies to show that 
all   Lagrangean torus orbits must project regularly to $M$. Furthermore, by Lemma (\ref{NOSL}) we know that the under the finite geometric multiplicity condition (\ref{HYP})   and uniform boundedness condition on the eigenfunctions, there do not exist any singular leaves of the moment map. Consequently,   the proof of Theorem (\ref{RM}) in the  case of Laplacians is a direct consequence of the following:

\begin{lem}\label{flatlem} Suppose that the geodesic flow $G^t$ of  $(M,g)$
commutes with a
Hamiltonian $\R^n$ action.  Suppose that there are no singular levels of the moment map,
and   suppose that each regular Lagrangean orbit $\R^n \cdot (x, \xi)$
has a non-singular projection to $M$.
Then $(M,g)$ is a flat manifold. \end{lem}

\noindent{\bf Proof}: We will give two proofs of the lemma.
\bigskip

\noindent{\bf First Proof}:
\bigskip

The first proof uses  Mane's theorem (\ref{MANE}):  Since the foliation
by orbits has no singular leaves, 
Mane's theorem implies that $(M,g)$ has no conjugate points.  Since each leaf is
compact,  it must be a torus
 which covers $M$.  Thus,  there exists a cover $p:T^n \to M$.  Lift the metric
to $p^*g$ on $T^n$.  The lifted metric must have no conjugate points since the
universal covering metric is the same.
By the Burago-Ivanov theorem (\ref{HOPF}),  the metric is
flat. \qed
\bigskip\

In the second proof, we do not use Mane's theorem, and directly relate the
condition on torus projections
to non-existence of conjugate points.

\bigskip

\noindent{\bf Second Proof:}
\bigskip

 As above, let $\pi: T^*M - 0 \rightarrow M$ denote
the natural projection and
 let $\pi_I = \pi |_{T_I}.$  Since each $\pi_I: T_I \rightarrow M$ is
non-singular,
and $dim T_I = dim M$, $\pi_I$ must be a covering map.

\subsubsection{Case 1: $M$ is a torus}

Let us first assume that $M$ is a torus, i.e. diffeomorphic to $\R^n/\Z^n$;
we make no assumptions on the metric. 

From the fact that $p_I$ is a covering map, 
it follows by a result of
Lalonde-Sikorav (\cite{LS}) that
the degree of $\pi_I: T_I \rightarrow M$ equals 1 for all $I$.
 Since $\pi_I$ is a diffeomorphism,
there are well-defined inverse maps
$$\pi_I^{-1}: M \rightarrow T_I$$
with $K(I) = 1.$  They define sections of $\pi: S^*M \rightarrow M$ and
hence are given by graphs
of 1-forms
$\alpha_I: M \rightarrow S^*M.$   Thus, $|\alpha_I (x)| \equiv 1$ where
$|\cdot|$ is the
co-metric.  We have $\pi_I^{-1 *} \alpha =
 \alpha_I^* \alpha = \alpha_I$ where
$\alpha$ is the canonical 1-form.   Since the tori $T_I$ are Lagrangean,
and since
$d\alpha = \omega$, the 1-forms
are closed,  i.e. $d\alpha_I = 0.$

Now let $p: \tilde{M} \rightarrow M$ denote the universal cover of $M$ and
let $\Z^n$ denote
the deck transformation group, with generators $\alpha_1, \dots, \alpha_n.$
The metric  $g$ lifts to a
$\Z^n$-periodic metric $\tilde{g}$ on $\tilde{M}$.  We note that the
corresponding
geodesic flow $\tilde{G}^t$ is also completely integrable.  Indeed,  the
cover $p$
induces the universal cover $p_1: T^*\tilde{M} \rightarrow T^*M$ whose deck
transformation
group we continue to denote by $\Z^n$.
Then $\tilde{G}^t$ commutes with the $T^n$-action on $T^*M-0$
  generated by the lifted action integrals
$\tilde{I}_j = p_1^* I_j.$  The invariant tori $T_I$ therefore lift to
$\tilde{G}^t$-invariant
 level sets $\tilde{T}_I$ of $(\tilde{I}_1, \dots, \tilde{I}_n).$

Furthermore, the 1-forms $\alpha_I$ lift to $\Z^n$-invariant closed 1-forms
$\tilde{\alpha_I}$ on $\tilde{M}.$
They are exact $\tilde{M}$ and hence have the form $d B_I$ for some
`potential' $B_I \in C^{\infty}(\tilde{M})$.   The gradient $\nabla B_I$ is
then a
$\Z^n$-invariant vector field on $\tilde{M}$.  Since $|d B_I| \equiv 1$ we
have
$|\nabla B_I| \equiv 1.$  We now claim that the integral curves of $\nabla
B_I$ are
lifts of geodesics on $T_I.$

To see this, we  recall that the generator $\Xi_H$ of the geodesic flow lies
tangent
 to each torus $T_I$.  Hence for each $I$ it projects from $T_I$ to a
non-singular vector field
 $\pi_{I *} \Xi_H = \Xi_I$ on $M$.   We have
$$\langle \nabla B_I, \Xi_I \rangle = dB_I (\Xi_I) = \alpha_I(\Xi_I) =
\langle \alpha|_{T_I},\Xi_H |_{T_I} \rangle = 1$$
since $\Xi_H$ is a contact vector field for $(S^*M, \alpha).$  Since
$|\nabla B_I| = 1$
it follows that $\nabla B_I = \Xi_I$.  This relation holds for the lifts to
$\tilde{M}$ and
hence the integral curves of $\nabla B_I$ are the lifts of the geodesics on
$T_I.$

We now claim that $g$ has no conjugate points, i.e. that each geodesic of
$\tilde{g}$ on
$\tilde{M}$ is length minimizing between each  two points on it.
 This follows by a well-known argument:
Let $\tilde{x}$ be any point of $\tilde{M}$, let $\tilde{v}  \in
S_{\tilde{x}} \tilde{M}.$
and let $\gamma_{\tilde{v}}$ be the geodesic  of $\tilde{g}$ in the
direction $\tilde{v}.$
To see that $\gamma_{\tilde{v}}$ is length minimizing between $\tilde{x}$
and any other
point $\gamma_{\tilde{v}}(t_o)$, we project it to $S^*M$.  The image lies in
one of
the (possibly singular) invariant tori $T_I$ and by the above,
$\gamma_{\tilde{v}}$ is an
integral curve of
$\nabla B_I.$   If it is not length minimizing to $\gamma_{\tilde{v}}(t_o)$,
then there exists
 $s_o < t_o$ and a second
geodesic $\alpha$ with $\alpha(0) = \tilde{x}, \alpha(s_o) =
\gamma_{\tilde{v}}(t_o).$
This leads to a contradiction since
$$B_I (\alpha(s_o)) = \int_0^{s_o} \langle \nabla B_I, \alpha'(s) \rangle ds
=
\int_0^{t_o} \langle \nabla B_I, \gamma_{\tilde{v}}'(s) \rangle ds = t_o >
s_o$$
but
$$t_o = |\int_0^{s_o} \langle \nabla B_I, \alpha'(s) \rangle ds| \leq s_o $$
as $|\nabla B_I| = 1$.   Therefore, $(T^n, g)$ is a torus without conjugate
points.  Theorem
A then follows in this case  from the recent proof by Burago-Ivanov
\cite{BI} of the Hopf conjecture
that a
metric on $T^n$ with no conjugate points is flat.\qed

\subsubsection{The general case} We now consider the general case where $M$
is only
covered by a torus $T^n$ (namely $T_I$ for each $I$).  We denote by $p: T^n
\to M$ a
fixed d-fold covering map. For notational clarity we denote the metric on
$M$ by $g_M$.
By Lemma (\ref{nonsing}), there is a Hamiltonian torus action on $T^*M-0$
with
the property that every orbit projects non-singularly to $M$.

Let $g_T = p^* g_M$ be the  metric induced on $T^n$ by the cover.  We claim
that
$g_T$ is a flat metric. Since $p : (T^n, g_T) \to (M, g)$ is a Riemannian
cover,
this will imply that $g_M$ is a flat metric and conclude the proof of (a).

To prove $g_T$ is flat, we  lift the torus foliation of $T^*M - 0$ to $T^* T^n -
0$. Given a metric $g$ on a manifold $X$ we denote by $\tilde{g}: TX \to T^*X$
the
induced bundle map  $\tilde{g}(X) = g(X, \cdot).$  We also consider the
bundle map:
$d p : T T^n \to TM$.  Since   $dp_x$ is a fiber-isomorphism for each $x \in
T^n$, $p$ is a d-fold covering map.  It
follows that
$$F : T^*(T^n) \to T^*M, \;\;\;\; F: = \tilde{g_M} d \rho \tilde{g}_T^{-1}$$
is also a d-fold covering map. Let $\mathcal T$ denote the foliation of
$T^*M-0$
by orbits of the torus action.  We define  $F^{-1} \mathcal T$ to be the
foliation
of $T^* T^n - 0$ whose leaves are given by $\tilde{T}_I:= F^{-1} T_I$ where
$\{T_I\}$ are the leaves
of $\mathcal T.$  (The associated 
involutive distribution
 of the $n$-planes $ \tilde{T}_{x, \nu} \subset T_{x, \nu} T^* T^n - 0$ is
defined by 
 $dF( \tilde{T}_{x, \nu}) = T_{F(x, \nu)} T_{I(F(x, \nu)}$.)  This foliation
could also defined as orbits of the commuting Hamiltonians $F^* I_j$ on $T^*
T^n-0$.  Each of the leaves is compact, hence a torus. 
We note that  $F: \tilde{T}_I \to T_I$ is always  a smooth covering map.  

  We then have the commutative  diagrams:
\begin{equation}\label{comm-diagram} \begin{array}{lcr} \tilde{T}_I
& \stackrel{F}{\rightarrow} & T_I \\ \pi \downarrow & & \downarrow \pi \\T^n
 &\stackrel{p}{\rightarrow}&M
\end{array}\end{equation}

We claim that the  map $\pi: \tilde{T}_I \to T^n$ is  non-singular.  If not,
the map $\pi \circ F: \tilde{T}_I \to M$
would be singular.  But  as observed above, it is a covering map.  
It further follows by the result of \cite{LS} that $\pi:  \tilde{T}_I \to T^n$
has degree one,
hence is a diffeomorphism. 

 We have now reduced to the previous case of the
torus: the metric
$g_T$ must be a flat metric, hence $g_M$ must be flat. This completes the second proof of Theorem (\ref{RM}) in the case of torus actions. \qed

\subsection{Proof of Theorem (\ref{RM}) for Schroedinger operators}

We now consider the case of semiclassical Schroedinger operators $\hbar^2 \Delta + V.$
Our proof in the homogeneous case (i.e. $V = 0$) was based on the use of semiclassical
pseudodifferential operators, so it generalizes with little change. 

\begin{proof}

 We fix an energy level $E$ and consider eigenvalues
of $\hbar^2 \Delta + V$ lying in $[E - C\hbar^{1 - \delta}, E + C \hbar^{1 - \delta}]$ for some fixed $C > 0.$ The eigenfunctions
we consider are the joint eigenfunctions of $P_{1}, \dots, P_{n}$ with joint eigenvalues
$(E_{j}(\hbar)=b_j^{(1)}(\hbar), \dots, b_j^{(n)}(\hbar))$ respectively,  satisfying $b_{j}^{(1)}(\hbar) \in [E - C\hbar^{1 - \delta}, E + C \hbar^{1 - \delta} ]$ for some $0 < \delta <1.$  We recall that $b=(b^{(1)}=E \, , \, b^{(2)},...,b^{(n)})$ and $E$ corresponds to the energy shell $X_{E}$  of the classical Hamiltonian $1/2 |\xi|^{2}_{g} + V$ corresponding to the quantum Hamiltonian $P_{1}= \hbar^{2} \Delta + V$. By assumption, the eigenfunctions corresponding to these joint eigenvalues are uniformly
bounded independently of $\hbar \leq \hbar_{0}.$

By Proposition (\ref{TB}), it follows that all Lagrangean torus orbits of $\Phi_t^E$ on $X_E$ project regularly
to the base. Indeed, the proof that the torus $\Lambda^{(j)}(b)$ projects regularly only
involves trace formula and quantum limits over joint eigenvalues in the set $\{(  E_{j}(\hbar)= b_j^{(1)}(\hbar) \, , b_{j}^{(2)}(\hbar) \, ,..., \, b_{j}^{(n)}(\hbar) \,) \, : \,  |b_j(\hbar) - b|
\leq \hbar^{1 - \delta}\}.$ . Hence our assumption on uniform boundedness of the  eigenfunctions of $P_{1}= \hbar^{2}\Delta +V$ with eigenvalues in the interval $[E-c \hbar^{1-\delta}, E+ c \hbar^{1-\delta}]$
is sufficient to obtain the result of Proposition (\ref{TB}) for the tori on the energy shell $X_E.$ 

  Hence, by a simple covering space argument, we can without loss of generality assume that  the base manifold  is a torus. By Lemma (\ref{NOSL}), there are no singular levels
of the moment map ${\mathcal P}|_{X_E}.$ Hence $X_E$ has a smooth Lagrangrean foliation
invariant under $\Phi_t^E.$
By Proposition (\ref{BP}),  we must have that $E > V_{max}$ and  the Jacobi metric $(E - V) g$ is flat.

If we additionally assume that the sup norms are bounded indepedently of $\hbar$ and
$E$ in some interval $[E_0 - \epsilon, E_0 + \epsilon]$, then the Jacobi metrics
$(E - V) g$ are flat for all $E$ in this interval, and it follows by Corollary (\ref{BPcor})
that $g$ is flat and $V$ is constant. \end{proof}

\section{Problems and Conjectures}

We conclude  with some problems conjectures on integrable systems and their eigenfunctions.

\subsection{Symplectic geometry of toric integrable systems}

Some of the ideas of this paper are relevant to purely geometric problems. 

\begin{conj}  Suppose that $g$ is a metric on $\R^n/\Z^n$ which
is toric
integrable.  Then $g$ is flat. \end{conj}

This would follow from the solution of the Hopf conjecture and from 

\begin{conj}  Up to symplectic equivalence, the only homogeneous Hamiltonian
torus action
on $T^*(\R^n / \Z^n)$ is the standard one ($\Phi_t(x, \xi) = (x + t\xi, \xi).$)
\end{conj}

Indeed, the geodesic flow of $(\R^n/\Z^n, g)$ would  preserve the Lagrangean
foliation defined by
orbits of $\Phi_t$ and hence by Mane's theorem $g$ would have no conjugate
points.

Since the time of the original submission of this article, these conjectures have been proved by E. Lerman
and N. Shirokova \cite{LS}.

\subsection{Eigenfunctions}

We assume throughout that the Laplacian or Schroedinger operator was quantum completely integrable. It is
natural to ask if the hypothesis can be weakened to classical integrability.

\begin{conj}\label{NOTQCI}  Suppose that $(M, g)$  is a compact Riemannian manifold with completely
integrable geodesic flow.  Suppose that $\Delta_g + V$ is a Schroedinger operator on $(M, g)$  all of whose  ONBE's
have uniformly bounded sup norms. Then $(M, g)$ is flat. \end{conj}

Without the assumption of quantum complete integrability,  it is not even  known whether  eigenfunctions localize on level sets of the classical
moment map. 

There are also interesting problems in the converse direction. We will explain the difficulty

of the next conjecture when we come to multiplicities.

\begin{conj} \label{GENERIC}
Suppose that $\Delta_g + V$ is a Schroedinger operator on a flat manifold  $(M, g)$. Then for generic $V$,  are  the
eigenfunctions uniformly bounded?  \end{conj}

We further note that all the questions about sup norms are equally reasonable in the

non-compact case.

\subsection{KAM and classically non-integrable systems}

We now consider the extent to which even classical complete integrability can be dropped. It is plausible that sup norm blow-up occurs whenever there exists
a stable elliptic orbit of the geodesic flow. In that case one can construct quasimodes
associated to the orbit which do blow up. The relation between modes and quasimodes can be
quite complicated in general, but it is plausible that there should exist a sequence of modes
which also blows up. KAM systems always contain such stable elliptic orbits. We plan to consider
these issues in a future article.

\subsection{Multiplicities and sup norms} There are (well-known) relations between eigenvalue

multiplicities and sup-norm blow up of eigenfunction.
If there exists a sequence of eigenvalues of unbounded multiplicity,
then there exists   an ONBE with unbounded  sup norms.
 Indeed,  for each $x$, consider the eigenfunction  $\Pi_E (x, \cdot)$
where $\Pi_E$ is   the orthogonal projection onto the eigenspace $V_E$.
Then $ \Pi_E (x, \dot)$ has $L^2$-norm equal to $\sqrt{\Pi_E(x,x)}$.  So the
normalized
eigenfunction is $\phi_E^x(\cdot):= \Pi_N(x,\cdot)/\sqrt{\Pi_N(x,x)}$. It is
well-known and easy to see (by the Schwartz inequality) that
$\phi_E^x(\cdot)$  has its maximum at $x$, where it equals
$\sqrt{\Pi_N(x,x)}$.  Since  $\int_M \Pi_N(x,x) dvol(x) = m(E)$ (with $dvol(x)$ 
the volume form), there must exist $x$ so that $\Pi_N(x,x) \geq
m(E).$ Hence $||\phi_E^x(\cdot)||_{\infty} \geq \sqrt{m(E)}.$ When $(M,
g)$ is a rational
torus,  $L^{\infty}(\lambda, M, g)$
therefore  grows at a polynomial rate while $\ell^{\infty}(\lambda, M, g)$  stays
bounded.

For instance,
on a flat torus $\R^n/L$, an
ONBE of the standard Laplacian $\Delta_0$ is given by the exponentials $e^{i \langle \lambda, x\rangle},$ with
$\lambda \in \Lambda := L^*$,
the dual lattice to $L$.  The associated eigenvalue is $E= |\lambda|^2$ and
its multiplicity $m(E)$ is the number
of lattice points of $\Lambda$ on the sphere of radius $\sqrt{E}$. Counting this number is
a well-known problem in
number theory when the lattice is rational.  When $L = \Z^n$, for instance, the
multiplicity
function $m(E|)$ has logarithmic
growth for $n = 2$, and polynomial growth in higher dimensions. 

Under a perturbation by a potential $\epsilon V$, there exists a smoothly varying
orthonormal basis of eigenfunctions (sometimes called the Kato-Rellich basis). It is
possible that for some potential $V$ on $\R^n / \Z^n$ , the Kato-Rellich basis
for the perturbation  $\Delta_0 + \epsilon V$   may be a smooth deformation
of the eigenfunctions just described with high sup norms. If so, it is then possible that
even if the multiplicity is broken and all eigenvalues become simple, the eigenfunctions
can still have unbounded sup norms.  Conjecture (\ref{GENERIC}) states that such potentials
should be sparse. It would be of some interest to understand if there exist any potentials
for which sup norm blow-up occurs.

The most extreme case of multiplicity is of course that on the standard
sphere $(S^2, g_0)$.  At this time of writing, it remains an open problem whether $\ell^{\infty}(\lambda, g) =
O(1)$ on the standard sphere. The best result to date is  the upper bound of VanderKam
\cite{V},  that for a `random' ONB of
eigenfunctions $\{\phi_{\lambda}\}$ the sup-norms
satisfy $||\phi_{\lambda}||_{\infty} / ||\phi_{\lambda}||_{L^2} =
O(\sqrt{\log \lambda}),$ i.e. $\ell^{\infty}(\lambda, S^2, g_0) = {\mathcal
O}(\sqrt{\log \lambda}).$
Our methods do not apply to this problem.

\subsection{Quantitative problems}

Can one  weaken  the hypothesis of uniform boundedness of eigenfunctions in
$L^{\infty}$  in the rigidity results?  It is plausible that our rigidity
results holds as long as   $L^{\infty}(\lambda, M, g)$  lies below
some threshold. One may ask the same question for the analogous   $L^p$
quantities $L^p(\lambda, M, g)$. 
In  \cite{TZ} (see also [T1][T2]), we analyse sup norm blow-up of eigenfunctions near singular levels
(among other things). We
also study some cases of sup norm blow up near singular projections of regular levels. To obtain
a threshfold of some generality one needs to estimate the minimal blow up corresponding
to the possible types of singular behaviour.

\end{document}